\documentclass[11pt,leqno]{article}
\usepackage{amsmath,amsthm,amssymb,amsxtra}
\usepackage[all]{xy}

\let\geq\geqslant
\let\leq\leqslant

\makeatother

\newcommand{\st}[1]{\ensuremath{^{\scriptstyle \textrm{#1}}}}

\newcommand\bigcheck[1]{#1 \raise1ex\hbox{$\hspace{-1ex}{}^\vee$}}
\newcommand\sucheck[1]{#1 \raise0.5ex\hbox{$\hspace{-1ex}{}^\vee$}}

\newcommand{\alphaparenlist}{
  \renewcommand{\theenumi}{\alph{enumi}}%
  \renewcommand{\labelenumi}{(\theenumi)}%
}

\newcommand{\romanparenlist}{
  \renewcommand{\theenumi}{\roman{enumi}}%
  \renewcommand{\labelenumi}{(\theenumi)}%
}

\newcommand{\Romanlist}{
  \renewcommand{\theenumi}{\Roman{enumi}}%
  \renewcommand{\labelenumi}{\theenumi.}%
}

\newcommand{\ann}{\mathop{\rm ann}\,}
\newcommand{\ad}{\mathop{\rm ad}\,}
\newcommand{\bmu}{\bar{\mu}}
\newcommand{\bnu}{\bar{\nu}}
\newcommand{\ch}{{\rm ch}}
\newcommand{\com}{{\rm com}}
\newcommand{\charge}{{\rm charge}}
\newcommand{\Cur}{{\rm Cur}}
\newcommand{\even}{\mathop{\rm even \, }}
\newcommand{\gh}{{\rm gh}}
\newcommand{\im}{\mathop{\rm im}\,}
\newcommand{\Image}{\mathop{\rm Image \, }}
\newcommand{\Lie}{\mathop{\rm Lie \, }}
\newcommand{\mult}{\mathop{\rm  mult \, }}
\renewcommand{\ne}{\mathop{\rm ne}\,}
\newcommand{\re}{\mathop{\rm re}\,}
\newcommand{\tw}{{\rm tw}}

\renewcommand{\det}{{\rm det}}

\newcommand{\End}{\mathop{\rm End }}
\renewcommand{\Im}{\mathop{\rm Im  \, }}

\newcommand{\odd}{{\rm odd}}

\newcommand{\Res}{\mathop{\rm Res  \, }}
\renewcommand{\sl}{s\ell}
\newcommand{\str}{{\rm str}}

\newcommand{\sdim}{\mathop{\rm sdim \, }}

\newcommand{\tr}{\rm tr \, }
\newcommand{\vac}{|0\rangle}

\newcommand{\A}{\mathcal{A}}
\newcommand{\C}{\mathcal{C}}

\newcommand{\CC}{\mathbb{C}}

\newcommand{\NN}{\mathbb{N}}

\newcommand{\RR}{\mathbb{R}}
\newcommand{\ZZ}{\mathbb{Z}}

\newcommand{\fg}{\mathfrak{g}}
\newcommand{\fh}{\mathfrak{h}}
\newcommand{\fhs}{\mathfrak{h}^{\natural}}

\newcommand{\fn}{\mathfrak{n}}

\renewcommand{\hat}{\widehat}

\makeatletter
\renewcommand\section{\@startsection {section}{1}{\z@}%
                                   {-3.5ex \@plus -1ex \@minus -.2ex}%
                                   {2.3ex \@plus.2ex}%
                                   {\normalfont\large\bfseries}}
\renewcommand\subsection{\@startsection{subsection}{2}{\z@}%
                                     {-3.25ex\@plus -1ex \@minus -.2ex}%
                                     {0ex \@plus .0ex}%
                                     {\normalfont\normalsize\bfseries}}

\@addtoreset{equation}{section}
\makeatother

\newtheorem{theorem}{Theorem}[section]

\newtheorem{corollary}{Corollary}[section]
\newtheorem{proposition}{Proposition}[section]
\newtheorem*{lemma*}{Lemma}

\theoremstyle{definition}

\theoremstyle{remark}
\newtheorem{remark}{Remark}[section]

\makeatletter
\newcounter{subremark}[remark]
\@addtoreset{subremark}{remark}
\renewcommand{\thesubremark}{\alph{subremark}}

\newenvironment{subremark}{
    \medskip
    \stepcounter{subremark}
    \noindent
    (\thesubremark)
  }%
{ 
  }
\makeatother

\newtheorem{example}{Example}[section]

\begin{document}

\title{Quantum reduction in the twisted case}

 \author{Victor G. Kac\thanks{Department of Mathematics, M.I.T.,
    Cambridge, MA 02139, USA.~~kac@math.mit.edu}~~\thanks{Supported in part by NSF grant
     DMS-0201017.}~~and Minoru Wakimoto\thanks{Graduate School
     of Mathematics, Kyushu University, Fukuoka 812-8581,
     Japan~~wakimoto@math.kyushu-u.ac.jp}~~\thanks{Supported by Grant-in-aid 13440012 for
 scientific research Japan.}}

\begin{abstract}
We study the quantum Hamiltonian reduction for affine 
superalgebras in the twisted case. This leads to a general representation 
theory of all superconformal algebras, including the twisted ones
(like the Ramond algebra). In particular, we find general free field
realizations and determinant formulae. 
\end{abstract}

\maketitle

\section{Introduction}
\label{sec:0}

This paper is a continuation of papers \cite{KRW} and \cite{KW} on
structure and representation theory of vertex algebras $W_k(\fg
,x)$ obtained by quantum Ham-iltonian reduction from the affine
superalgebra $\hat{\fg}$.  The datum one begins with
is a quadruple $(\fg ,x,f,k)$, where $\fg$ is a simple
finite-dimensional Lie superalgebra with a non-zero invariant
even supersymmetric bilinear form $(\, . \, | \, . \, )$, $x$ is
an element of $\fg$ such that $\ad \, x$ is diagonalizable with
eigenvalues in $\frac{1}{2}\ZZ$, $f$ is an even element of $\fg$
such that $[x,f]=-f$ and the eigenvalues of $\ad \, x$ on the
centralizer $\fg^f$ of $f$ in $\fg$ are non-positive, and $k \in
\CC$.  Recall that a pair $\{ x,f \}$ satisfying the above
properties can be obtained from an $\sl_2$-triple $\{ e,x,f \}$,
so that $[x,e]=e$, $[x,f]=-f$, $[e,f]=x$.

We associate to the quadruple $(\fg ,x,f,k)$ a homology complex
$\C (\fg ,x,k)= (V_k (\fg) \otimes F_{\ch}\otimes F_{\ne} \, , \,
d_0)$, where $V_k (\fg)$ is the universal affine vertex algebra
of level $k$ associated to the affine superalgebra $\hat{\fg}$,
$F_{\ch}$ is the vertex algebra of free charged superfermions
based on $\fg_+ +\fg^*_+$ with reversed parity, $F_{\ne}$ is
the vertex algebra of free neutral superfermions based on
$\fg_{1/2}$, and $d_0$ is an explicitly constructed odd
derivation of the vertex algebra $\C (\fg ,x,k)$ whose square is
$0$.  Here $\fg_+$ (resp. $\fg_{1/2}$) denotes the sum of
eigenspaces of $\ad \, x$ with positive eigenvalues (resp. with
eigenvalue $1/2$), and we drop $f$ from the notation since its
different choices are conjugate.  The vertex algebra $W_k (\fg
,x)$ is the homology of the complex $(\C (\fg ,x,k),\, d_0)$.

In the present paper we begin with a diagonalizable automorphism
$\sigma$ of $\fg$ with modulus 1 eigenvalues, which leaves  invariant
the bilinear form $( . \, | \, . \, )$ and keeps the elements
$x$ and $f$ fixed.  The automorphism $\sigma$ gives rise to the
twisted affine superalgebra $\hat{\fg}^{\tw}$ and the
corresponding twisted vertex algebra $V_k (\fg ,\sigma)$, and to
the twisted vertex algebras $F^{\tw}_{\ch}$ and $F^{\tw}_{\ne}$,
and we consider the twisted complex
\begin{displaymath}
  (\C (\fg ,\sigma ,x,k) =V_k (\fg ,\sigma) \otimes
     F^{\tw}_{\ch} \otimes F^{\tw}_{\ne}\, ,\, d^{\tw}_0 )\, .
\end{displaymath}
Its homology is the twisted vertex algebra $W_k (\fg ,\sigma
,x)$, which is the main object of our study.

In the case when $\sigma =1$ we recover the ``Neveu-Schwarz
sector'' $W_k (\fg ,x)$ studied in our previous papers \cite{KRW}
and \cite{KW} (and earlier in \cite{FF,FKW,BT,FB,ST}, and many other
works, see \cite{BS}.)  In the case when $\sigma =\sigma_R$,
where $\sigma_R |_{\fg_{\bar{0}}} =1$ and $\sigma_R
|_{\fg_{\bar{1}}} =-1$, we obtain the ``Ramond sector''.  This
terminology comes  from the fact that taking the smallest simple
Lie superalgebra $\fg = osp (1|2)$ and the only possible choice
of $x$, we obtain as $W_k (\fg ,x)$ the vertex algebra associated
to the usual Neveu-Schwarz  algebra, and as $W_k (\fg ,\sigma_R
,x)$ the twisted vertex algebra associated to the usual Ramond
algebra.
Likewise, taking $\fg =\sl (2|1)$, $\sl (2|2))/\CC I$, $osp
(3|2)$ or $D(2,1 ;a)$, and $x$ the suitable multiple of the
highest root $\theta$ of one of the simple components of
$\fg_{\bar{0}}$, all possible choices of $\sigma$ produce all
possible twists of the $N=2$, $N=4$, $N=3$ and big $N=4$
superconformal algebras.

This leads us to a unified representation theory of all twisted
superconformal algebras, in particular to  unified free field
realizations and determinant formulas.

As in \cite{FKW} and \cite{KRW}, we construct also a functor $M
\mapsto H(M)$ from the category of restricted
$\hat{\fg}^{\tw}$-modules to the category of  $\ZZ$-graded $W_k (\fg ,\sigma
,x)$-modules and compute the Euler-Poincar\'e character of $H(M)$
in terms of the character of $M$.  In a forthcoming paper
\cite{KW4} we shall develop a theory of characters of $W_k (\fg
,\sigma ,x)$ using this functor.

\section{An overview of twisted formal distributions}
\label{sec:1}

Let $R$ be a Lie conformal superalgebra.  Recall that this is a
$\ZZ /2\ZZ$-graded $\CC [\partial]$-module, endowed with
$j$\st{th} products denoted by $a_{(j)}b$, $j \in \ZZ_+$,
satisfying certain axioms \cite{K4}.  One associates to $R$ a Lie
superalgebra
\begin{equation}
  \label{eq:1.1}
  \Lie (R) = R \, [t,t^{-1}]/\Image (\partial \otimes 1 + 1 \otimes
  \partial_t)\, ,
\end{equation}
endowed with the following bracket, where $a_{(\mu)}$ stands for $a
\otimes t^{\mu} \in R \, [t,t^{-1}]=R \otimes \CC \, [t,t^{-1}]$:
\begin{equation}
  \label{eq:1.2}
  [a_{(\mu)}, b_{(\nu)}]=\sum_{j \in \ZZ_+} \binom{\mu}{j}
     (a_{(j)}b)_{(\mu +\nu-j)}\, .
\end{equation}
Introducing formal distributions
\begin{equation}
  \label{eq:1.3}
  a(z) = \sum_{\mu \in \ZZ} a_{(\mu)} z^{-\mu -1}\, ,
        \quad a \in R \, ,
\end{equation}
one rewrites (\ref{eq:1.2}) as
\begin{equation}
  \label{eq:1.4}
  [a(z) \, , \, b(w)] =\sum_{j \in \ZZ_+} (a_{(j)}b)
     (w) \partial^j_w \delta (z-w)/j ! \, .
\end{equation}
One also has:
\begin{equation}
  \label{eq:1.5}
  (\partial a) (z) =\partial_z a(z) \, .
\end{equation}
(The fact that Lie $R$ is a Lie superalgebra and the
distributions $\{ a(z) \}_{a \in R}$ form a local system is encoded in
the axioms of $R$.)

Let now $\sigma$ be a diagonalizable automorphism of $R$.  We shall
always assume for simplicity that all eigenvalues of $\sigma$
have modulus~$1$. We have:
\begin{equation}
  \label{eq:1.6}
  R=\bigoplus_{\bar{\mu}\in \RR /\ZZ} R^{\bar{\mu}}, \hbox{ where }
     R^{\bar{\mu}} =\{ a \in R | \sigma (a) = e^{2\pi i\bar{\mu}}a
        \} \, .
\end{equation}

Here and further $\bar{\mu}$ denotes the coset $\mu + \ZZ$ of
$\mu \in \RR$.  We associate to the pair $(R ,\sigma)$ the
$\sigma$-\emph{twisted} Lie superalgebra
\begin{equation}
  \label{eq:1.7}
  \Lie (R ,\sigma) =\bigoplus_{\mu \in \RR} \quad
    (R^{\bmu} \otimes t^{\mu})/\Image (\partial \otimes 1 +
        1\otimes \partial_t) \, ,
\end{equation}
endowed with bracket (\ref{eq:1.2}) (except that now $\mu$ and
$\nu$ are not necessarily integers).  Denoting by $a_{(\mu)}$ the
image of $a\otimes t^{\mu}$ in $\Lie (R,\sigma)$, and introducing
the twisted formal distributions
\begin{equation}
  \label{eq:1.8}
  a^{\tw}(z) =\sum_{\mu \in \bmu} a_{(\mu)} z^{-\mu -1}\, ,
      \quad a \in R^{\bmu} \, ,
\end{equation}
we get the twisted analogue of (\ref{eq:1.4}):
\begin{equation}
  \label{eq:1.9}
  [a^{\tw}(z) \, , \, b^{\tw}(w)] =\sum_{j \in \ZZ_+}
     (a_{(j)}b)^{\tw}(w) \partial^j_w \delta_{\bmu}(z-w)/j!\, ,
\end{equation}
where
\begin{equation*}
  \delta_{\bmu} (z-w)=z^{-1}\sum_{\mu \in \bmu}
     \left( \frac{w}{z} \right)^{\mu}
\end{equation*}
is the twisted formal $\delta$-function.

Assuming that the $\CC[\partial]$-module $R$ is generated by a finite set $Q$,
introduce a descending filtration of the Lie superalgebra $L=\Lie
(R,\sigma)$ by subspaces $F_jL = \{ a_{(\mu)} | a \in Q , \mu \geq
  j \}$, and define a completion $U(L)^{\com}$ of its universal
  enveloping algebra $U(L)$, which consists of all series $\sum_i
  u_i$ such that for each $N \in \RR$ all but finitely many of
  the $u_i$'s lie in $U(L) (F_NL)$.  The automorphism $\sigma$ of
  $R$ induces one of $L$ and of $U(L)^{\com}$ in the obvious way,
  which we again denote by $\sigma$.

A $U(L)^{\com}$-valued \emph{twisted formal distribution} is an expression of
the form
\begin{displaymath}
  a(z) =\sum_{\mu \in \bmu} a_{(\mu)}z^{-\mu -1} \, ,
\end{displaymath}
where $\bmu = \mu + \ZZ$, $\sigma (a_{(\mu)}) = e^{2\pi i \bmu}
a_{(\mu)}$ and $a_{(\mu)} \in U(L)^{\com}$ satisfy the property
that for each $N \in \RR$, $a_{(\mu)} \in U(L)(F_NL)$ for $\mu
\gg 0$ and all the $a_{(\mu)}$ have the same parity, denoted by
$p(a) \in \ZZ /2\ZZ$.  It is clear that the derivative $\partial_z
a(z)$ of a twisted formal distribution $a(z)$ is also a twisted formal distribution.

In order to define a normally ordered product of twisted formal distributions
$a(z)$ and $b(z)$, we need to define a splitting $a(z) = a(z)_+ +
a(z)_-$ into creation and annihilation parts $a(z)_+$ and
$a(z)_-$.  For that we choose $s_a$ in the coset $\bmu$ and let
\begin{equation}
  \label{eq:1.10}
  a(z)_+ =\sum_{\mu <s_a} a_{(\mu)} z^{-\mu -1} \, , \,
  a(z)_- =\sum_{\mu \geq s_a} a_{(\mu)} z^{-\mu-1}\, .
\end{equation}
If $a(z)$ is a non-twisted formal distribution, i.e.,~$\bmu =\ZZ$, then one
may choose $s_a =0$, so that $\partial_z (a(z)_{\pm}) = (\partial_z
a(z))_{\pm}$, but for twisted formal distributions such a choice is
impossible.  After making a choice of $s_a$, one defines the normally
ordered product of twisted formal distributions in the usual
way:
\begin{displaymath}
  :a(z) b(z) : = a(z)_+ b(z) + (-1)^{p(a)p(b)}
     b(z) a(z)_- \, .
\end{displaymath}
It is easy to see that this is again a $U(L)^{\com}$-valued formal distribution.
As usual, one defines the normally ordered product of
more than two formal distributions from right to left, e.g.~$:abc
: = : a:bc::$.

Denote by $V(R)$ the subspace of $U (\Lie R)^{\com}$ consisting
of all normally ordered products of formal distributions
(\ref{eq:1.3}) and $1$.  This is one of the constructions of the
universal enveloping vertex algebra of the Lie conformal algebra
$R$ \cite{KRW} (cf. \cite{K4},\cite{GMS},\cite{BK}). The infinitesimal
translation operator $\partial$ of $V(R)$ is defined by
(\ref{eq:1.5}).  The $j$\st{th} product $a_{(j)}b$ on $V(R)$ is
defined by (\ref{eq:1.4}) for $j \in \ZZ_+$, and by $a_{(-j-1)} b=
:(\partial^ja)b:/j!$ for $j \in \ZZ_+$. The automorphism $\sigma$ of $R$
induces an automorphism of $V(R)$, and we have its eigenspace decomposition:
\begin{displaymath}
  V(R) =\bigoplus_{\bmu \in \RR /\ZZ} V^{\bmu}(R) \, .
\end{displaymath}
Likewise, denote by $V(R, \sigma)$ the subspace of $U(\Lie
(R,\sigma))^{\com} [[z,z^{-1}]]$ consisting of all normally
ordered products of twisted formal distributions (\ref{eq:1.8})
and $1$.  This is called a $\sigma$-\emph{twist} of the vertex
algebra $V(R)$.  (It is independent of the choices of $s_a$ used
in the definition of normally ordered products.)  The subspace
$V(R,\sigma)$ is $\sigma$-invariant, so that we have the
decomposition into its eigenspaces:
\begin{displaymath}
  V(R,\sigma) =\bigoplus_{\bmu \in \RR /\ZZ} V^{\bmu}(R ,\sigma) \, .
\end{displaymath}
%

The following result is well known.

\begin{proposition}
  \label{prop:1.1}
The map $a(z) \mapsto a^{\tw}(z)$ ($a\in R^{\bmu}, \bmu \in \RR/\ZZ$)
extends uniquely to a $\sigma$-eigenspace preserving vector space isomorphism
$V(R) \to V (R,\sigma)$, $a(z) \mapsto a^{\tw}(z)$,
%
%
satisfying the following properties $(a\in V^{\bmu}(R),b \in V(R))$:
\begin{eqnarray}
  \label{eq:1.11}
  1^{\tw}&=&1\, ,\\
  \label{eq:1.12}
  (\partial a)^{\tw}(z) &=& \partial_z a^{\tw}(z) \, , \\
  \label{eq:1.13}
  [a^{\tw}(z), b^{\tw}(w)] &=& \sum_{j \in \ZZ_+}
     (a_{(j)}b)^{\tw}(w)\partial^j_w \delta_{\bmu}(z-w)/j!\, , \\
  \label{eq:1.14}
  :a^{\tw}(z) b^{\tw} (z) : &=& \sum_{j \in \ZZ_+}
  \binom{s_a}{j}  (a_{(j-1)}b)^{\tw} (z) z^{-j}\, .
\end{eqnarray}

\end{proposition}

A module $M$ over the filtered Lie superalgebra $L = \Lie (R,
\sigma)$ is called \emph{restricted} if any vector of $M$ is
annihilated by some $F_jL$.  Such an $L$-module can be uniquely
extended to a module over the associative algebra $U(L)^{\com}$.
Restricting this module to $V(R,\sigma)$, we obtain, in view of
Proposition~\ref{prop:1.1}, what is called a
$\sigma$-\emph{twisted} module $M$ over the vertex algebra $V(R)$.

In the examples of Lie conformal superalgebras $R$ that follow we
use the $\lambda$-bracket $[a_{\lambda}b] =\sum_{j \in \ZZ_+}
\tfrac{\lambda^j}{j!} a_{(j)}b$.  Due to sesquilinearity
$([\partial a_{\lambda}b] =-\lambda [a_{\lambda}b]$,
$[a_{\lambda}\partial b] =(\partial +\lambda) [a_{\lambda}b])$,
the $\lambda$-brackets of generators of the $\CC [\partial]$-module
$R$ determine the $\lambda$-bracket on $R$.  Recall also that an
element $K$ of $R$ is called central if $[K_{\lambda}R] =0=[R_{\lambda}K]$.

\begin{example}{(twisted currents and Sugawara construction)}.
  \label{ex:1.1}
Let $\fg$ be a simple finite-dimensional Lie superalgebra with a
non-degenerate supersymmetric invariant bilinear form $(\, . \, |
\, . \, )$.  The associated Lie conformal superalgebra is
\begin{displaymath}
  \Cur \fg = (\CC [\partial]\otimes \fg) \oplus \CC K \, ,
\end{displaymath}
where $K$ is a central element and
\begin{displaymath}
  [a_{\lambda}b]=[a,b]+\lambda (a|b)K \, , \quad
    a,b \in 1 \otimes \fg \equiv \fg \, .
\end{displaymath}

\end{example}

Given a complex number $k$, denote by $V_k(\fg)$ the quotient of
the universal enveloping vertex algebra $V (\Cur \fg)$ by the
ideal generated by $K-k$.  This is called the universal affine
vertex algebra of level $k$.

Let $\sigma$ be a diagonalizable automorphism of the Lie
superalgebra $\fg$, keeping the bilinear form $(\, . \, | \, . \,
)$ invariant.  It extends to an automorphism of $\Cur \fg$, also
denoted by $\sigma$, by letting $\sigma (P (\partial)\otimes
a)=P(\partial)\otimes \sigma (a)$, $\sigma (K)=K$.  Let $\fg
=\oplus_{\bmu \in \RR /\ZZ} \fg^{\bmu}$, where $\fg^{\bmu}=\{ a
\in \fg |\,  \sigma (a) =e^{2\pi i \bmu}a \}$, be the eigenspace
decomposition of $\fg$ for $\sigma$.  Then the corresponding
$\sigma$-twisted Lie superalgebra $\Lie (\Cur \fg,\sigma)$ is a
twisted Kac-Moody affinization
\begin{displaymath}
  \hat{\fg}^{\tw '} =\bigoplus_{\mu \in \RR} (\fg^{\bmu}\otimes
  t^{\mu}) \oplus \CC K \, ,
\end{displaymath}
with the bracket $(a \in \fg^{\bmu}, b \in \fg^{\bnu})$:
\begin{displaymath}
[at^{\mu},bt^{\nu}] =[a,b] t^{\mu + \nu} + \mu(a|b)
\delta_{\mu ,-\nu} K \, , \quad [K,\hat{\fg}^{\tw '}]=0 \, .
\end{displaymath}
The formal distributions
\begin{displaymath}
  a^{\tw}(z) =\sum_{\mu \in \bmu} (at^{\mu}) z^{-\mu -1}\, ,
  \quad a \in \fg^{\bmu}\, , 
\end{displaymath}
are called \emph{twisted currents}.  They generate (by taking
derivatives and normally ordered products) the $\sigma$-twist
$V(\Cur \fg,\sigma)$ of the vertex algebra\break$V(\Cur \fg)$.  As in
the non-twisted case, denote by $V_k (\fg , \sigma)$ the
quotient by the ideal generated by $K-k$; this is the $\sigma$-twist
 of the vertex algebra $V_k (\fg)$.

Choosing dual bases $\{ a_i \}$ and $\{ a^i \}$ of $\fg$,
compatible with the eigenspace decomposition for $\sigma$, so
that $(a_i | a^j) = \delta_{ij}$, define the twisted Sugawara
field in $V_k (\fg ,\sigma)$ (assuming that $k+h\spcheck \neq 0$):
\begin{displaymath}
  L^{\fg ,\tw}(z) = \frac{1}{2(k+h\spcheck)} \sum_i (-1)^{p(a_i)}
     : a_i a^i :^{\tw}(z) \, .
\end{displaymath}
Writing $L^{\fg,\tw}(z)=\sum_{n \in \ZZ} L^{\fg,\tw}_n z^{-n-2}$, and using
the non-twisted Sugawara construction and formula
(\ref{eq:1.13}), we obtain that the $L^{\fg,\tw}_n$ satisfy the
relations of the Virasoro algebra with central charge
 $ c(k) = k\sdim \fg/(k+h\spcheck)$.
Using formula (\ref{eq:1.14}), we can rewrite $L^{\fg,\tw}(z)$ in
terms of twisted currents
and numbers $s_i =s_{a_i}$ (see (\ref{eq:1.10})):
\begin{eqnarray}
  \label{eq:1.15}
\lefteqn{\hspace{2ex}
L^{\fg,\tw} (z) = \frac{1}{2(k+h\spcheck)} \left( \sum_i (-1)^{p(a_i)}:
    a^{\tw}_i (z) a^{i,\tw}(z) : \right.  }  \\
 \nonumber  \qquad && \left. -\sum_i (-1)^{p(a_i)} s_i [a_i,a^i]^{\tw}(z) z^{-1}-k
    \sum_i (-1)^{p(a_i)} \binom{s_i}{2} z^{-2} \right)  \, .
\end{eqnarray}

\begin{example}{(twisted neutral free superfermions)}.
  \label{ex:1.2}
Let $A$ be a finite-\break dimensional vector superspace with a
non-degenerate skew-supersymmetric bilinear form $\langle \, . \, , \,
. \, \rangle$.  The associated Clifford Lie conformal
superalgebra is
\begin{displaymath}
  C(A) = (\CC [\partial]\otimes A) \oplus \CC K \, ,
\end{displaymath}
where $K$ is a central element and
\begin{displaymath}
  [a_{\lambda} b] = \langle a,b \rangle K \, .
\end{displaymath}
Denote by $F(A)$ the quotient of the universal enveloping vertex
algebra of $C(A)$ by the ideal generated by $K-1$.
\end{example}

Let $\sigma$ be a diagonalizable automorphism of the space $A$,
keeping the bilinear form $\langle \, . \, , \, . \, \rangle$
invariant.  As above, it extends to an automorphism $\sigma$ of
the Lie conformal superalgebra $C(A)$.  Let $A=\oplus_{\bmu \in
  \RR /\ZZ}$ $A^{\bmu}$ be the eigenspace decomposition for
$\sigma$.  Then the corresponding $\sigma$-twisted Lie
superalgebra $\Lie (C(A), \sigma)$ is a twisted Clifford
affinization
\begin{displaymath}
  \hat{A}^{\tw} =\oplus_{\mu \in \RR} (A^{\bmu} \otimes t^{\mu})
     \oplus \CC K_A
\end{displaymath}
with the bracket
\begin{displaymath}
  [at^{\mu}\, , \, bt^{\nu}]=\langle a,b \rangle \delta_{\mu,-\nu -1}
      K_A \, , \quad [K_A \, , \, \hat{A}^{\tw}]=0 \, .
\end{displaymath}
We shall work in the Clifford algebra
$U(\hat{A}^{\tw})^{\com}/(K-1)$.  The formal distributions
\begin{displaymath}
  \Phi^{\tw} (z) = \sum_{\mu \in \bmu} (\Phi t^{\mu})
  z^{-\mu -1} \, , \quad \Phi \in A^{\bmu} \, ,
\end{displaymath}
are called twisted neutral free superfermions.  They generate the
$\sigma$-twist $F(A,\sigma)$ of the vertex algebra $F(A)$.

Choosing dual bases $\{ \Phi_i \}$ and $\{ \Phi^i \}$ of $A$,
compatible with the eigenspace decomposition for $\sigma$, we
let
\begin{displaymath}
  L^{\ne,\tw} (z) =\frac{1}{2} \sum_i (-1)^{p(\Phi_i)}:
     \Phi_i \partial \Phi^i :^{\tw}(z) \, .
\end{displaymath}
Writing $L^{\ne,\tw}(z) =\sum_{n\in\ZZ} L^{\ne,\tw}_n z^{-n-2}$, we
obtain a Virasoro algebra with central charge $c=-\tfrac{1}{2}
\sdim A$.  As in the previous example, using formula
(\ref{eq:1.14}), we obtain:
\begin{eqnarray}
  \label{eq:1.16}
L^{\ne,\tw}(z) &=& \frac{1}{2}\sum_i (-1)^{p(\Phi_i)}:
  \Phi^{\tw}_i (z) \partial \Phi^{i,\tw}(z) : \\
\nonumber
&& -\frac{1}{2} \sum_i (-1)^{p(\Phi_i)} \binom{s_i}{2}z^{-2} \, .
\end{eqnarray}

\begin{example}{(twisted charged free superfermions)}.
  \label{ex:1.3}
In notation of Example~\ref{ex:1.3}, assume that $A=A_+ \oplus
A_-$, where both $A_+$ and $A_-$ are isotropic and
$\sigma$-invariant subspaces.  Choose a basis $\varphi_i$ of
$A_+$, compatible with the eigenspace decomposition of $A_+$ for
$\sigma$, and its dual basis $\varphi^*_i$ of $A_-$, so that
$\langle \varphi_i,\varphi^*_j \rangle = \delta_{ij}$, and define
\emph{charge} by
\begin{equation}
  \label{eq:1.17}
  \charge (\varphi_i)=1 \, ; \quad \charge (\varphi^*_i) =-1\, .
\end{equation}

\end{example}

The formal distributions $\varphi^{\tw}_i (z)$ and
$\varphi^{*\tw}_i (z)$ are called twisted  charged free
superfermions.  Relation~(\ref{eq:1.17}) gives rise to the charge
decomposition:
\begin{equation}
  \label{eq:1.18}
  F(A \, , \, \sigma) =\oplus_{m \in \ZZ}
     F_m (A\, , \, \sigma)\,.
\end{equation}
For a collection of complex numbers $(m_j) \in \CC^{\dim A_+}$ we can
define a Virasoro formal distribution
\begin{displaymath}
  L^{\ch,\tw} (z) =-\sum_i m_i : \varphi^*_i \partial\varphi_i
     :^{\tw} (z) + \sum_i (1-m_i):\partial \varphi^*_i\varphi_i :^{\tw}(z)
\end{displaymath}
with central charge $\sum_i (-1)^{p(\varphi_i)} (12 m^2_i -12m_i
+2)$.  Using formula~(\ref{eq:1.14}), we obtain
\begin{eqnarray}
  \label{eq:1.19}
\lefteqn{\hspace{1ex}  L^{\ch,\tw}(z) = -\sum_i m_i : \varphi^{*\tw}_i(z)
     \partial\varphi^{\tw}_i (z) : } \\
\nonumber\quad\quad &&+ \sum_i (1-m_i):\partial\varphi^{*\tw}_i (z)
   \varphi^{\tw}_i (z) :+\sum_i (-1)^{p(\varphi_i)}
   \binom{s_i}{2}z^{-2}\, .
\end{eqnarray}

\section{The twisted complex}
\label{sec:2}

Let $\fg$ be a simple finite-dimensional Lie superalgebra with a
non-degenerate even supersymmetric invariant bilinear form $(\, . \, |
\, . \, )$.  Fix an even element $x$ of $\fg$ such that $\ad x$
is diagonalizable with half-integer eigenvalues, and let
\begin{equation}
  \label{eq:2.1}
  \fg = \oplus_{j \in \frac{1}{2}\ZZ} \fg_j
\end{equation}
be the eigenspace decomposition.  Let
\begin{displaymath}
  \fg_+ = \oplus_{j>0} \fg_j \, , \,
  \fg_- =\oplus_{j<0} \fg_j \, , \,
  \fg_{\leq} = \fg_0 \oplus \fg_- \, .
\end{displaymath}
An even element $f \in \fg_{-1}$ is called \emph{good} if its
centralizer $\fg^f$ in $\fg$ lies in $\fg_{\leq}$, and the
gradation (\ref{eq:2.1}) is called \emph{good} if it admits a good
element.  We shall assume that the grading (\ref{eq:2.1}) is good
and we shall fix a good element $f \in \fg_{-1}$ (all good
elements form a Zariski dense orbit of the group $\exp
\fg_{0,\even}$, hence nothing depends on the choice of $f$).

The most interesting good gradings come from $\sl_2$-triples $\{
e,x,f \}$, where $[x,e]=e$, $[x,f]=-f$, $[e,f]=x$, which are
called Dynkin gradings.  However, there are many other good
gradings.  In the Lie algebra case they are classified in
\cite{EK}.

An important role is played by the following bilinear form
$\langle \, . \, , \, . \, \rangle_{\ne}$ on $\fg_{1/2}$:
\begin{equation}
  \label{eq:2.2}
  \langle a,b \rangle_{\ne}=(f|[a,b])\, ,
\end{equation}
which is skew-supersymmetric, even and non-degenerate.

Fix an automorphism $\sigma$ of $\fg$ with the following three
properties:

\romanparenlist
\begin{enumerate}
\item 
$\sigma (x) =x$, $\sigma (f) =f$;

\item 
$(\sigma (a)|\sigma (b)) = (a|b)$ for all $a,b \in \fg$;

\item 
$\sigma$ is diagonalizable and all its eigenvalues have modulus~$1$.
\end{enumerate}

We shall construct a twisted vertex algebra $W_k (\fg ,\sigma
, x)$ depending on a complex parameter $k$.  For $\sigma =1$
this coincides with the vertex algebra $W_k (\fg ,x,f)$ studied
in \cite{KRW} and \cite{KW} (we shall drop $f$ from the notation,
since different choices of $f$ give isomorphic algebras).

Introduce the following $\tfrac{1}{2}\ZZ$-graded subalgebra of
$\fg$:
\begin{equation}
  \label{eq:2.3}
  \fg (\sigma) =\oplus_{j \in \frac{1}{2}\ZZ} \fg_{j}(\sigma)\, ,
  \hbox{ where } \fg_j (\sigma) = \{ a \in \fg_j |\sigma (a)
     =(-1)^{2j}a \} \, .
\end{equation}

Choose a $\sigma$-invariant Cartan subalgebra $\fh$ of the even
part of $\fg_0$, and choose a triangular decomposition of $\fg
(\sigma)$, compatible with the gradation (\ref{eq:2.3}):
\begin{equation}
  \label{eq:2.4}
  \fg (\sigma) = \fn (\sigma)_- \oplus \fh^{\sigma}
     \oplus \fn (\sigma)_+ \, ,
\end{equation}
where $\fh^{\sigma}$ denotes the fixed point set of $\sigma$ on
$\fh$, such that the following properties hold:

\romanparenlist
\begin{enumerate}
\item 
$\fn (\sigma)_{\pm}$ are isotropic with respect to $(\, . \, | \,
. \, )$ nilpotent subalgebras normalized by $\fh^{\sigma}$,

\item 
$f \in \fn (\sigma)_+$,

\item 
$\fn_{1/2}(\sigma)_+ :=\fg_{1/2}(\sigma)\cap \fn (\sigma)_+$ is a
maximal isotropic subspace of $\fg_{1/2}(\sigma)$ with respect to
$\langle \, . \, , \, . \, \rangle_{\ne}$,

\item 
$\fn_{1/2} (\sigma)_-$ is a direct sum of a maximal isotropic
subspace $\fn_{1/2}(\sigma)_-^{'}$ of  $\fg_{1/2}(\sigma)$ with respect to $\langle \,
. \, , \, . \, \rangle_{\ne}$ and at most $1$-dimensional subspace
$\fg^0_{1/2}(\sigma)$, normalized by $\fh^{\sigma}$.
\end{enumerate}

Here and further we let $\fn_j (\sigma)_{\pm}=\fn (\sigma)_{\pm}
\cap \fg_j (\sigma)$. We thus have the following decomposition:
\begin{equation}
\label{eq:2.5}
\fg_{1/2} (\sigma)= \fn_{1/2} (\sigma)_+ + \fg_{1/2}(\sigma)_0
    + \fn_{1/2}(\sigma)_-^{'} \, ,
\end{equation}
where $    \epsilon (\sigma) :=\dim \fg_{1/2}(\sigma)_0 \leq 1$.
Note that $\epsilon (\sigma ) \neq 0$ iff $\dim \fg_{1/2}
(\sigma)$  is odd.

\begin{remark}
  \label{rem:2.1}
Let $\fg_0 (\sigma)^f$ be the centralizer in $\fg_0 (\sigma)$ of
$f \in \fg_{-1}(\sigma)$, and assume that there exists a
semisimple element $h_0$ in $\fg_0 (\sigma)^f$ such that all
eigenvalues of $\ad h_0$ on $\fg_{1/2}(\sigma)$ are real numbers
and the multiplicity of zero is at most $1$
(it follows from \cite{EK}, Theorem~1.5, that such an $h_0$ with
all eigenvalues non-zero exists if $\fg (\sigma)$ is a Lie
algebra).  Let $m$ denote the minimal absolute value of the
non-zero eigenvalues.  Let $H_0 \in \fh^{\sigma}$ be a regular
element of $\fg (\sigma)$ such that all eigenvalues of $\ad H_0$
are real, the eigenvalue on $f$ is positive and their absolute
values are smaller than $m$.  Let $\fn (\sigma)_+$ (resp.~$\fn
(\sigma)_-$) denote the span of the eigenvectors of $\ad
(h_0+H_0)$ in $\fg (\sigma)$ with positive (resp. negative)
eigenvalues.  This gives us a decomposition (\ref{eq:2.4})
satisfying all properties (i)---(iv).  It is because the bilinear
form $(\, . \, | \, . \, )$ is non-degenerate on $\fg (\sigma)$
and the bilinear form $\langle \, . \, , \, . \, \rangle_{\ne}$ is
non-degenerate and invariant on $\fg (\sigma)_{1/2}$ with respect
to $\fg_0 (\sigma)^f$.

\end{remark}

Let $D=-L^{\fg ,\tw}_0$.  Recall that we have $(a \in
\fg^{\bar{\mu}})$:
\begin{displaymath}
  [D,at^{\mu}] =\mu (at^{\mu}) \, , \, [D,K]=0\, .
\end{displaymath}
As usual, we shall consider the extension of the Kac-Moody
affinization (see Example~\ref{eq:1.1}):
\begin{displaymath}
  \hat{\fg}^{\tw}=\CC D \ltimes \hat{\fg}^{\tw\prime}\, .
\end{displaymath}

The decomposition (\ref{eq:2.4}) induces a triangular
decomposition of the Lie superalgebra $\hat{\fg}^{\tw}$ (see
Example~\ref{ex:1.1}):
\begin{eqnarray}
  \label{eq:2.6}
  \hat{\fg}^{\tw} &=& \hat{\fn}_- \oplus \hat{\fh}\oplus
      \hat{\fn}_+ \, , \\
\noalign{\nonumber{where}}\\
    \label{eq:2.7}
      \hat{\fh} &=& \fh^{\sigma} + \CC K +\CC D\, , \\[1ex]
   \label{eq:2.8}
      \hat{\fn}_+ &=& \sum_{j \in \frac{1}{2}\ZZ}
          (\fn_j (\sigma)_+\otimes t^{-j} +
          \sum_{\substack{\mu \in \RR\\j+\mu >0}} \fg^{\bmu}_j
          \otimes t^{\mu})\, , \\
   \label{eq:2.9}
      \hat{\fn}_- &=& \sum_{j \in \frac{1}{2} \ZZ}(\fn_j(\sigma)_-
         \otimes t^{-j} + \sum_{\substack{\mu \in \RR \\
             j+\mu<0}}
            \fg^{\bmu}_j \otimes t^{\mu}) \, .
\end{eqnarray}

As usual, we extend the (non-degenerate) invariant bilinear for
$(\, . \, | \, . \, )$ from $\fh^{\sigma}$ to $\hat{\fh}$ by:
\begin{displaymath}
  (\CC K + \CC D |\fh^{\sigma})=0 \, , \, (K|K) = (D|D)=0\, , \,
  (K|D)=1\, .
\end{displaymath}
This bilinear form is non-degenerate, and we shall identify
$\hat{\fh}$ and $\hat{\fh}^*$ via this form.

Denote by $A_{\ne}$ the vector superspace $\fg_{1/2}$ with the
bilinear form $ \langle \, . \, , \, . \, \rangle_{\ne}$.  Denote
by $A_+$ (resp.~$A_-$) the superspace $\fg_+$ (resp. its dual
$\fg^*_+$) with reversed parity, and let $A_{\ch} =A_+ \oplus
A_-$.  Let $ \langle \, . \, , \, . \, \rangle_{\ch}$ be the
skew-supersymmetric bilinear form on $A_{\ch}$ defined by
\begin{displaymath}
  \langle A_{\pm} \, , \, A_{\pm}\rangle_{\ch}=0 \, , \,
  \langle a \, , \, b^* \rangle_{\ch}=b^* (a) \hbox{ for }
   a \in A_+ \, , \quad b^* \in A_- \, .
\end{displaymath}
The automorphism $\sigma$ of $\fg$ induces automorphisms of
$A_{\ne}$ and $A_{\ch}$, which we again denote by $\sigma$, that
preserve the respective bilinear forms.  Finally, fix a complex
number $k$ such that $k+h\spcheck \neq 0$.

We shall associate to the data $(\fg ,x,f,k,\sigma)$ a twisted
differential vertex algebra
\begin{displaymath}
  (\C (\fg,\sigma ,x,k), \, d^{\tw}_0)\, .
\end{displaymath}

Consider the twisted Kac-Moody affinization $\hat{\fg}^{\tw\prime}$ and
Clifford affinizations $\hat{A}^{\tw}_{\ne}$ and $\hat{A}^{\tw}_{\ch}$
(see Examples~\ref{ex:1.1}, \ref{ex:1.2} and \ref{ex:1.3}).  Let
$L$ be the direct sum of these Lie superalgebras with filtration
$F_N L=F_N \hat{\fg}^{\tw\prime} + F_N \hat{A}^{\tw}_{\ne} + F_N
\hat{A}^{\tw}_{\ch}$.  Let $U(L)^\com$ be the completed via this
filtration universal enveloping algebra of $L$ and let
$U_k(L)^{\com}$ be the quotient of $U(L)^{\com}$ by the ideal
generated by $K-k$, $K_{A_{\ne}}-1$, $K_{A_{\ch}}-1$.  Recall that
twisted currents, twisted neutral superfermions and twisted
charged superfermions generate (via taking derivatives and
normally ordered products) the twisted vertex algebras $V_k (\fg
,\sigma)$, $F(A_{\ne} ,\sigma)$ and $F(A_{\ch} ,\sigma)$.  We denote
by $F(\fg,\sigma ,x)$ the twisted vertex algebra generated by the
last two and by $\C (\fg ,\sigma ,x,k)$ the one generated by all
three types of formal distributions.  We have:
\begin{displaymath}
  F(\fg ,\sigma ,x) = F (A_{\ch} ,\sigma) \otimes
      F(A_{\ne} ,\sigma)\, ,
   \C(\fg ,\sigma ,x,k) = V_k (\fg ,\sigma) \otimes F(\fg ,\sigma ,x)\, .
\end{displaymath}

By letting $\charge (V_k (\fg ,\sigma))=\charge (F(A_{\ne} ,\sigma))=0$
and using (\ref{eq:1.17}), one has the induced charge
decompositions:
\begin{displaymath}
  F(\fg ,\sigma ,x)=\oplus_{m \in \ZZ} F(\fg ,\sigma ,x)^\tw_m
       \, , \, \C (\fg ,\sigma ,x,k)=\oplus_{m \in \ZZ}
       \C^\tw_m \, .
\end{displaymath}
\vspace{-3ex}

In order to define the differential $d^{\tw}_0$, and for further
use, choose a basis 
$\{ u_i \}_{i \in S}$ of $\fg$ compatible with the gradation (\ref{eq:1.1}),
the $\sigma$-eigenspace decomposition and the root space decomposition
with respect to $\fh^{\sigma}$. 
A part of this basis is a basis of $\fg_m$ ($m\in \frac{1}{2}\ZZ$),
and of $\fg_+$. As in \cite{KW}, we denote the corresponding subsets of indices of $S$
by $S_m$ and $S_+$ respectively.
Define the structure
constants $c^{\ell}_{ij}$ by $[u_i,u_j]=\sum_{\ell} c^{\ell}_{ij}
u_{\ell}$.  Denote by $\{ \varphi_i \}_{i \in S_+}$, $\{
\varphi^*_i \}_{i \in S_+}$ the corresponding basis of $A_+$ and its
dual basis of $A_-$, so that $\langle \varphi_i , \varphi^*_j
\rangle_{\ch} =\delta_{ij}$, and by $\{ \Phi_i \}_{i \in
S_{1/2}}$ the corresponding basis of $A_{\ne}$.
We shall denote by $\{u^i\}$ the dual basis of $\fg$ with respect
to the form $(.|.)$ and by $\{\Phi^i\}_{i \in S_{1/2}}$ the dual
basis of $A_{\ne}$ with respect to the form $\langle.,.\rangle_{\ne}$.

Recall that in the non-twisted case, i.e.,~when $\sigma =1$, we
defined $d_0 =\Res_z \, d(z)$, where $d(z)$ is the following
formal distribution of the vertex algebra $\C (\fg ,1,x,k)$
\cite{KRW}, \cite{KW}:
\begin{eqnarray*}
  d(z) &=& \sum_{i \in S_+}(-1)^{p(u_i)} u_i (z)
            \otimes \varphi^*_i (z) \otimes 1\\
         &&- \frac{1}{2}\sum_{i,j,\ell \in S_+}
            (-1)^{p(u_i)p(u_{\ell})} c^{\ell}_{ij}\otimes :
            \varphi_{\ell} (z)\varphi^*_i(z)\varphi^*_j (z) :\otimes 1 \\
         && + \sum_{i \in S_+} (f|u_i)\otimes \varphi^*_i
            (z) \otimes 1 + \sum_{i \in S_{1/2}} 1 \otimes
            \varphi^*_i (z) \otimes \Phi_i (z) \, .
\end{eqnarray*}
Further on, for simplicity of notation, we shall omit the tensor
sign.  Note that in each summand of $d(z)$, factors are commuting
formal distributions.  Hence the corresponding (via
Proposition~\ref{prop:1.1}) twisted formal distribution $d^{\tw}(z)$ of
$\C (\fg ,\sigma ,x,k)$ is given by the same expression as
$d(z)$, where all factors $u_i (z)$, $\varphi^*_i (z)$, etc. are
replaced by $u^{\tw}_i (z)$, $\varphi^{*\tw}_i (z)$, etc. (it is
because $:a^{\tw}(z) b^{\tw}(z):=(a_{(-1)}b)^{\tw}(z)$ if
$a_{(j)}b=0$ for $j \in \ZZ_+$, by (\ref{eq:1.14})).  Since
$[d(z) ,d(w)]=0$, it follows that $[d^{\tw}(z),d^{\tw}(w)]=0$.
Hence the odd element $d^{\tw}_0=\Res_z \, d^{tw}(z)$ has the
property that $(d^{\tw}_0)^ 2 =0$.  Note also that $d^{\tw}_0$ is a
derivation of all products of the twisted vertex algebra $\C (\fg
,\sigma ,x,k)$ and $d^{\tw}_0 (\C^{\tw}_m) \subset
\C^{\tw}_{m-1}$.

Denote the homology of the complex $(\C (\fg ,\sigma ,x,k), \,
d^{\tw}_0)$ by $W_k (\fg ,\sigma ,x)$.
This $\sigma$-twisted vertex algebra is called the \emph{$\sigma$-twisted quantum reduction
for the triple} $(\fg,\sigma,x)$.
The automorphism $\sigma$
of $\fg$ obviously induces a diagonalizable automorphism of the
vertex algebra $\C (\fg ,1,x,k)$, commuting with the operator
$d_0$.  Hence it induces a diagonalizable automorphism, also
denoted by $\sigma$, of the vertex algebra $W_k (\fg ,1,x)$.

The most important formal distribution of $W_k (\fg ,\sigma ,x)$
is the $\sigma$-twist $L^{\tw}(z)$ of the Virasoro formal distribution $L(z)$, defined by~(2.2)  
of \cite{KW}:
\begin{displaymath}  
L^{\tw}(z) =L^{\fg ,\tw} (z) +L^{\ne,\tw}(z) +  L^{\ch,\tw}(z) +\partial_z x^{\tw}(z)\, ,
\end{displaymath}
where the $m_i$ in (\ref{eq:1.19}) are defined by $u_i \in \fg_{m_i}$.  

Recall that the building blocks of the vertex algebra $W_k (\fg
,x)$ are the following formal distributions \cite{KW}:
\begin{displaymath}
  J^{(v)} (z)=  v(z) +\sum_{i,j \in S_+} (-1)^{p(u_i)}
     c_{ij} (v) : \varphi_i (z) \varphi^*_j (z) : \, ,
\end{displaymath}
where $v \in \fg$ and $c_{ij}(v)$ is the matrix of $\ad v$ in the
basis $\{ u_i \}$, i.e.,~$[v,u_j]=\sum_i c_{ij} (v) u_i$.  Using
(\ref{eq:1.14}), we obtain the following formula for the
corresponding twisted formal distribution ($v \in \fg$):
\begin{eqnarray}
  \label{eq:2.10}
  J^{(v)\tw}(z) &=& v^{tw} (z) +\sum_{i,j \in S_+}
      (-1)^{p(u_i)} c_{ij}(v) : \varphi^{\tw}_i (z)
         \varphi^{*\tw}_j (z) : \\
         \nonumber
         &&- \sum_{i \in S_+} (-1)^{p(u_i)} s_i c_{ii}
         (v) z^{-1} \, .
\end{eqnarray}

Theorem 4.1 of \cite{KW} implies the following result.

\begin{theorem}
  \label{th:2.1}

\alphaparenlist
\begin{enumerate}

\item 
For each $a \in (\fg^{\bmu}_{-j})^f$, $j \geq 0$, there exists a
$d^{\tw}_0$-closed twisted formal distribution $J^{\{ a \} ,\tw}
(z)$ in $\C (\fg ,\sigma ,x,k)$ of conformal weight $1+j$ (with
respect to $L^{\tw}(z)$) such that $J^{\{ a \}
  ,\tw}(z)-J^{(a),\tw}(z)$ is a linear combination of normally
ordered products of the twisted formal distributions
$J^{(b)\tw}(z)$, where $b \in \fg_{-s}$, $0 \leq s<j$, the
twisted formal distributions $\Phi^{\tw}_i (z)$, where $i \in
S_{1/2}$, and their derivatives.

\item 
The homology classes of the formal distributions $J^{\{ a_i \}
  ,\tw} (z)$, where $\{ a_i \}$ is a basis of $\fg^f_{\leq 0}$
compatible with the $\frac{1}{2} \ZZ$-gradation and
$\sigma$-eigenspace decomposition, strongly and freely generate $W_k (\fg
,\sigma ,x)$.

\item 
$W_k (\fg ,\sigma ,x)$ is a $\sigma$-twist of the vertex algebra
$W_k(\fg ,1,x )$.

\item 
$W_k (\fg ,\sigma ,x)$ coincides with the $0$\st{th} homology of
the complex $(\C (\fg ,\sigma ,x,k)$, $d^{\tw}_0)$.
\end{enumerate}
\end{theorem}

\section{Modules over $W_k (\fg ,\sigma ,x)$}
\label{sec:3}

Denote by $S'\subset S$ the subset of indices of the part of the basis  
$\{ u_i \}_{i \in S}$ of $\fg$, which is a basis of $\fg \mod \fh^{\sigma}$,
and let $S_0'=S_0\cap S'$ .
In the case when $\fh=\fh^{\sigma}$, $S'$ can be identified with the set of roots
of $\fg$ with respect $\fh$, but it is larger otherwise.

Recall that, given a diagonalizable automorphism $\sigma$ of a
vertex algebra $V$, so that $V=\oplus_{\bmu \in \RR /\ZZ}$ $
V^{\bmu}$ is its eigenspace decomposition, a $\sigma$-twisted
module $M$ over $V$ is a linear map $a \to a^{M,\tw}(z) = \sum_{n
\in \bmu} a^M_{(n)} z^{-n-1}$ $(a \in V^{\bmu})$ satisfying
equations (\ref{eq:1.12})---(\ref{eq:1.15}), where $a^M_{{(n)}}
\in \End M$ and for any $v \in M$, $a^M_{(n)} v =0$ if $n \gg
0$.  In other words, the collection of fields $a^{M,\tw}(z)$
forms a $\sigma$-twist of the vertex algebra $V$.  In this
section we shall discuss the properties of $\sigma$-twisted
modules over $W_k (\fg ,x)$ ($=$ modules over $W_k (\fg ,\sigma
,x)$) obtained by the $\sigma$-twisted quantum reduction from
restricted $\hat{\fg}^{\tw}$-modules.

We shall embed $\fh^{\sigma*}$ in $\hat{\fh}^*$ by letting
$\lambda \in \fh^ {\sigma *}$ be zero on $K$, and we define
$\delta \in \hat{\fh}^*$ by $\delta |_{\fh^{\sigma} + \CC K} =0$,
$\delta (D) =1$.  Recall that, given a triangular decomposition
(\ref{eq:2.6}), a \emph{highest weight module} over the Lie
superalgebra $\hat{\fg}^{\tw}$ of level $k$ and with highest
weight $\Lambda \in \fh^*$ is a $\hat{\fg}^{\tw}$-module $M$
which admits a non-zero vector $v_{\hat{\Lambda}}$, where
$\hat{\Lambda} =\Lambda +kD$, with the properties:

\romanparenlist
\begin{enumerate}
\item 
 $hv_{\hat{\Lambda}} =\hat{\Lambda} (h) v_{\hat{\Lambda}}$, $h \in \hat{\fh}$,

\item 
$\hat{\fn}_+ v_{\hat{\Lambda}}=0$,

\item 
$U(\hat{\fn}_-)v_{\hat{\Lambda}}=M$.
\end{enumerate}
For this reason, in the definition~(\ref{eq:1.10}) of the annihilation
part of the twisted current $u_i(z) $  ($i \in S'$), we choose
\begin{equation}
  \label{eq:3.1}
  s_{u_i} =\min \{ n |\, u_i \otimes t^n \hbox{ is non-zero and lies in }
      \hat{\fn}_+ \} \, , s_{h}=1\, \hbox{for}\, h\in \fh^{\sigma}.
\end{equation}
Since each summand $\fg_j$ of the gradation (\ref{eq:2.1}) is
$\sigma$-invariant, we have its $\sigma$-eigenspace
decomposition:
\begin{displaymath}
   \fg_j = \oplus_{\bmu \in \RR /\ZZ}
\fg^{\bmu}_j, \hbox{where}\, \fg^{\bmu}_j =\{ a \in \fg_j| \sigma (a)
=e^{2\pi i \bmu}a \}.
\end{displaymath}
Hence for a basis element $u_i \in
\fg^{\bmu_i}_{m_i}$ we can rewrite formula (\ref{eq:3.1}) for
$s_i =s_{u_i}$ ($i \in S'$) as follows:
\begin{eqnarray}
  \label{eq:3.2}
  s_i = \left\{
    \begin{array}{ll}
      \min \{ n \in \bmu_i |\,  n>-m_i \} \hbox{ if }
          & u_i \not\in \fn (\sigma)_+ \, , \\
       -m_i \hbox{ if } u_i \in \fn (\sigma)_+ \, .
    \end{array}\right.
\end{eqnarray}
It is easy to see that for a dual basis element $u^i \in
\fg^{-\bmu}_{-m_i}$ we have for $s^i =s_{u^i}$:
\begin{equation}
  \label{eq:3.3}
  s^i = 1-s_i \hbox{ for all }i \in S' \, .
\end{equation}

We extend this definition of annihilation operators to
$\hat{A}^{\tw}_{\ne}$ and $\hat{A}^{\tw}_{\ch}$ as follows:
\begin{equation}
  \label{eq:3.4}
  s_{\Phi_i} = s_i  \,(i \in S_{1/2}) \, , \, s_{\varphi_i}=s_i \, , \,
     s_{\varphi^*_i}=1-s_i \,\, (i \in S_+)\, .
\end{equation}
It is easy to see that we have
\begin{equation}
  \label{eq:3.5}
  s_{\Phi_i}=\mp 1/2  \hbox{ if }\Phi_i \in \fn_{1/2}
     (\sigma)_{\pm}\, , \, |s_{\Phi_i}|<1/2\hbox{ otherwise.}
\end{equation}
\begin{equation}
  \label{eq:3.6}
  s_{\Phi_i} + s_{\Phi^i} = \delta_{i,i_0}\, , \, \hbox{where}\,\, \langle \Phi_{i_0},\Phi_{i_0}
 \rangle_{\ne}  \neq 0.
\end{equation}

We write the generating fields in the form:
\begin{eqnarray*}
  u_i (z) &=& \sum_{n \in \bmu_i} u_{i,n}z^{-n-1}\, , \,
  \Phi_i (z) = \sum_{n \in \bmu_i +1/2} \Phi_{i,n}z^{-n-1/2}\, , \, \\
  \varphi_i (z) &=& \sum_{n \in \bmu_i} \varphi_{i,n}
     z^{-n-1}\, , \, \varphi^*_i (z) =\sum_{n \in -\bmu_i}
     \varphi^*_{i,n} z^{-n} \, .
\end{eqnarray*}

Each of the Clifford affinizations $\hat{A}^{\tw}_{\ne}$ and
$\hat{A}^{\tw}_{\ch}$ has a unique irreducible module, denoted by
$F^{\tw}_{\ne}$ and $F^{\tw}_{\ch}$, respectively, admitting a
non-zero vector $\vac_{\ne}$ and $\vac_{\ch}$, respectively,
killed by all annihilation operators:
\label{eq:3.7}
\begin{equation}
  \Phi_{i,n}\vac_{\ne} = 0 \hbox{ for } n \geq s_{i}+1/2 \, , \\
\end{equation}
\begin{equation}
\label{eq:3.8}
  \varphi_{i,n}\vac_{\ch} = 0 \hbox{ for }n \geq s_i \, , \,
  \varphi^*_{i,n} \vac =0 \hbox{ for } n\geq 1-s_i \, .
\end{equation}

Since these modules are restricted, they extend to the modules
over $F (A_{\ne} ,\sigma)$ and $F(A_{ch} ,\sigma)$ ($=$ twisted
modules over the vertex algebras $F(A_{\ne} ,\break$$ 1)$ and
$F(A_{\ch},1)$), respectively), hence
\begin{displaymath}
  F^{\tw} = F^{\tw}_{\ne} \otimes F^{\tw}_{\ch}
\end{displaymath}
is a module over $F (\fg ,\sigma ,x)$ ($=$ twisted module over the tensor product of these
vector algebras, $F(\fg ,1,x)$).  We let
\begin{displaymath}
  \vac = \vac_{\ne} \otimes \vac_{\ch} \in F^{\tw}\, .
\end{displaymath}

Thus, given a restricted $\hat{\fg}^{\tw}$-module $M$ with $K=kI$, we
extend it to a module over $V_k (\fg ,\sigma)$ ($=$ twisted
module over the vertex algebra $V_k (\fg ,1)$), then $M \otimes
F^{\tw}$ becomes a module over $\C(\fg,\sigma,x,k)$ (=twisted module over
the vertex algebra $\C (\fg,x,k)$). Passing to the homology of the complex
$\C^{\tw} (M)= (M \otimes F^{\tw},
d_0^{\tw})$, we obtain a $W_k(\fg,\sigma,x)$-module (=twisted $W_k(\fg,x)$-module)
$H^{\tw}(M)$. One has the charge decomposition of $\C^{\tw}(M)$ induced by that
of $F(\fg,\sigma,x)$ by setting the charge of $M$ to be zero. This induces a
decomposition as $W_k(\fg,\sigma,x)$-modules:
$H^{\tw}(M)=\sum_{j\in \ZZ} H_j^{\tw}(M)$.

Let $\Delta^{\sigma} \subset \fh^{\sigma *}$ be the set of
non-zero roots  of $\fg$ with respect to $\fh^{\sigma}$, counted
with their multiplicities.  
We may identify $\Delta^{\sigma}$ with a subset of $S'$,
which indexes root vectors attached to non-zero roots.
(Then the remaining elements of $S'$ index a basis of $\fh \mod \fh^{\sigma}$.)
Given one of the above basis root vectors $e_{\alpha}$, 
attached to $\alpha \in \Delta^{\sigma}$, we let $s_{\alpha}=s_{e_{\alpha}}$.  One
should keep in mind that the $s_{\alpha}$ corresponding to root vectors with  the
same $\alpha$ may be different (in the case $\fh^{\sigma} \neq \fh$).  

Recall that the set of roots $\hat{\Delta} \subset \hat{\fh}^*$
of the twisted affine Lie superalgebra $\hat{\fg}^{\tw}$ is
$\hat{\Delta } = \hat{\Delta}^{\re} \cup \hat{\Delta}^{\im}$,
where:
\begin{displaymath}
  \hat{\Delta}^{\re} = \{ \alpha + (m+s_{\alpha}) \delta | \, m
  \in \ZZ \, , \, \alpha \in \Delta^{\sigma} \} \, , \,
  \hat{\Delta}^{\im}=\{ m\delta | \, m \in E_0 \backslash \{ 0 \}
  \} \, ,
\end{displaymath}
where $E_0 = \{ \mu \in \RR
| \, e^{2\pi i \mu}$ is an eigenvalue of $\sigma$ on $\fh \}$,
and the roots are considered with their multiplicities.  Then we
have a subset $\hat{\Delta}_+ =\hat{\Delta}^{\re}_+ \cup
\hat{\Delta}^{\im}_+$ of positive roots in $\hat{\Delta}$,
corresponding to $\hat{\fn}_+$ (see (\ref{eq:2.8})), where
\begin{displaymath}
  \hat{\Delta}^{\re}_+ = \{ \alpha + (m+s_{\alpha}) \delta
    | \, m \in \ZZ_+ \, , \, \alpha \in \Delta^{\sigma} \}
    \, , \, \hat{\Delta}^{\im}_+ = \{ m\delta |\, m \in E_0\, ,\,
    m > 0 \} \, .
\end{displaymath}
%

Introduce the following subset of $\hat{\Delta}^{\re}_+$:
\begin{displaymath}
  \hat{\Delta}^{\re}_{++} = \{ \alpha + (m+s_{\alpha}) \delta |
  \, \alpha \in \Delta^{\sigma} \, , \,
  \alpha (x) \geq 0 \, , \, m \in \ZZ_+ \} \, .
\end{displaymath}

\begin{proposition}
  \label{prop:3.1}
(a) If $M$ is a restricted
$\hat{\fg}^{\tw}$-module and $v \in M$ is a
singular vector, i.e.,~$\hat{\fn}^{\tw}_+ v=0$, then
\begin{displaymath}
  d^{\tw}_0 (v \otimes \vac) =0 \, .
\end{displaymath}
(b) If $M$ is a Verma module over $\hat{\fg}^{\tw}$ with the highest
weight vector $v_{\hat{\Lambda}}$ and $v \in M$ is a
singular vector with highest weight $\hat{\Lambda}-n\alpha$, where
$\alpha \in \hat{\Delta}^{\re}_{++}$,
then the homology class of $v\otimes \vac$ in $H_0(M)$ is non-zero.
\end{proposition}

\begin{proof}
  We have:  $d^{\tw}_0 =A+B+C+D$, where
  \begin{eqnarray*}
    A &=& \sum_{i \in S_+}
          \sum_{\substack{p \in \bmu_i\\q\in -\bmu_i\\p+q=0}}
          (-1)^{p(u_i)} u_{i,p} \varphi^*_{i,q} \, , \\[1ex]
    B &=& -\frac{1}{2} \sum_{i,j,k \in S_+}
          \sum_{\substack{p \in \bmu_k\\q\in -\bmu_i\\
              r \in -\bmu_j\\p+q+r=0}}
          (-1)^{p(u_i)p(u_k)} c^k_{ij}\varphi_{k,p}
          \varphi^*_{i,q} \varphi^*_{j,r} \, , \\[1ex]
    C &=& \sum_{i \in S_+} (f|u_i)\varphi^*_{i,1}\, , \,
    D =  \sum_{i \in S_{1/2}}
          \sum_{\substack{p \in -\bmu_i\\q\in \bmu_i+1/2 \\p+q=0}}
          \varphi^*_{i,p} \Phi_{i,q+1/2}\, .
  \end{eqnarray*}

A summand of $A$ does not annihilate $v \otimes \vac$ only if
$p \leq s_i -1$, $q \leq s_i$, hence there are no such summands
since $p+q=0$.

A summand of $B$ does not annihilate $v \otimes \vac$ only if
$p \leq s_k-1$, $q \leq -s_i$, $r\leq s_j$, which happens only if
$p+q+r \leq s_k-s_i-s_j-1 \leq -1$, since $s_k \leq s_i+s_j$ when
$c^k_{ij}\neq 0$.  Hence there are no such summands.

If $(f|u_i) \neq 0$, then $(f t|u_it^{-1})\neq 0$, and since $ft
\in \hat{\fn}_+$, we obtain that $u_it^{-1} \in \fn_-$ and
therefore $s_i \geq 0$, by definition of $s_i$.  Hence
$\varphi^*_{i,1}$ is an annihilation operator (see
(\ref{eq:3.8})) and $C (v \otimes \vac )=0$.

Finally, if a summand of $D$ does not annihilate the $v \otimes \vac$, then
$p \leq s_i$ and $q+1/2 \leq s_i -1/2$ and therefore $p+q=-1$,
which is impossible since $p+q=0$.

This proves (a).  The proof of (b) is the same as in the
non-twisted case, see  \cite{KW}, Lemma~7.3.

\end{proof}

Next, we study the formal distribution $L^{\tw}(z)$ of $W_k
(\fg ,\sigma ,x)$.  
Using formulas (\ref{eq:1.15}), (\ref{eq:1.16}) and
(\ref{eq:1.19}) for the first
three summands, we obtain an
explicit expression for $L^{\tw}(z) =\sum_{n \in \ZZ}L^{\tw}_n
z^{-n-2}$. 
Note that the $L^{\tw}_n$ form a Virasoro
algebra with the same central charge as in the non-twisted case.
Examples~\ref{ex:1.1},
\ref{ex:1.2} and \ref{ex:1.3} give the following important formulas.

\begin{proposition}
  \label{prop:3.2}
Introducing the constants
\begin{eqnarray}
  \label{eq:3.9}
  s_{\fg} &=& -\frac{k}{2(k+h\spcheck)}\sum_{\alpha \in S'}
               (-1)^{p(\alpha)} \binom{s_{\alpha}}{2}\, , \\[1ex]
  \label{eq:3.10}
  s_{\ne} &=& \frac{1}{8}\epsilon (\sigma)
              -\frac{1}{2}\sum_{\alpha \in S_{1/2}} (-1)^{p(\alpha)}
                \binom{s_{\alpha}}{2}\, , \\[1ex]
\label{eq:3.11}
  s_{\ch} &=& \sum_{\alpha \in S_+} (-1)^{p(\alpha)}
             \left( \binom{s_{\alpha}}{2} +m_{\alpha} s_{\alpha} \right) \, ,
\end{eqnarray}
we have
\begin{eqnarray*}
  L^{\fg ,\tw}_0 &=& \frac{1}{2(k+h\spcheck)} \left( \sum_i
                    h_ih^i -\sum_{i \in S'}
                    (-1)^{p(\alpha)}s_{\alpha} \alpha
                    \right) +s_{\fg}+\ann  \, ; \\
  L^{\ne,\tw}_0 &=& s_{\ne}+\ann \vac_{\ne} \, , \,
  L^{\ch,\tw}_0 =s_{\ch}+\ann \vac_{\ch} \, ; \\[1ex]
  L^{\tw}_0 &=& L^{\fg ,\tw}_0 + L^{\ne,\tw}_0 +
  L^{\ch,\tw}_0 -x \, ,
\end{eqnarray*}
where $\ann $ (resp.~$\ann \vac$) denotes the sum of terms
which annihilate any singular vector in a $\hat{\fg}^{\tw}$-module $M$
of level~$k$ (resp. annihilate the vacuum vector), and $\{h_i\}$ and $\{h^i\}$
are dual bases of $\fh^{\sigma}$.
\end{proposition}


\begin{proof}
  We have:  $\sum_{\alpha \in S'} (-1)^{p(\alpha)} s_{\alpha}
  [u_{\alpha},u^{\alpha}] =\sum_i a_ih_i $, where $a_i \in \CC$.
  Hence $a_i= \sum_{\alpha \in S'} (-1)^{p(\alpha)} s_{\alpha}
  ([u_{\alpha}, u^{\alpha}] |\, h^i) =\sum_{\alpha \in S'}
  (-1)^{p(\alpha)} s_{\alpha} \alpha (h^i)$.  Hence $\sum_{\alpha
  \in S'}$\break$ (-1)^{p(\alpha)} s_{\alpha}
[u_{\alpha},u^{\alpha}]=\sum_{\alpha \in S'} (-1)^{p(\alpha)}
s_{\alpha} \alpha$.  The rest of the calculation is straightforward.
\end{proof}

\begin{corollary}
  \label{cor:3.1}
Let $v$ be a singular vector of a
$\hat{\fg}^{\tw}$
-module $M$
of level $k$ such that $av = \Lambda (a) v$, $a \in \fh^{\sigma}$, for
some $\Lambda \in \fh^{\sigma *}$.  Then $L^{\tw}_0 (v \otimes \vac) =h$ $v \otimes \vac$, where
\begin{displaymath}
  h=\frac{1}{2(k+h\spcheck)} ((\Lambda | \Lambda)-
      \sum_{\alpha \in S'}(-1)^{p(\alpha)} s_{\alpha} (\Lambda | \alpha)) -\Lambda(x)+ s_{\fg} +
      s_{\ne} + s_{\ch}\, .
\end{displaymath}

\end{corollary}
\begin{corollary}
  \label{cor:3.2}
Let $\gamma' =\frac{1}{2}\sum_{\alpha \in S'}
(-1)^{p(\alpha)}\alpha \in \fh^{\sigma *}$, and let
$\hat{\rho}^{\tw}$ be the Weyl vector rho for
$\hat{\fg}^{\tw}$
(i.e.,~($\hat{\rho}^{\tw}|\alpha_i)=\frac{1}{2}(\alpha_i |\alpha_i)$
for all simple roots $\alpha_i$ of $\hat{\fg}^{\tw}$).  Then
$\hat{\rho}^{\tw}|_{\fh^{\sigma}}=-\gamma'$.

\end{corollary}

\begin{proof}
  By Proposition~\ref{prop:3.2} we have in any highest weight
  $\hat{\fg}^{\tw}$-module of level $k$ with highest weight
  $\Lambda \in \fh^{\sigma *}$:
  \begin{equation}
    \label{eq:3.12}
    L^{\fg ,\tw}_0 v_{\Lambda}=\frac{1}{2(k+h\spcheck)}
    \left( (\Lambda |\Lambda )-2(\Lambda
      |\gamma')\right)+c_1 \, ,
  \end{equation}
where $c_1 \in \CC$ is independent of $\Lambda$.  On the other
hand, the operator $L^{\fg ,\tw}_0 +D$ commutes with
$\hat{\fg}^{\tw}$, hence equals $c_2 \Omega^{\tw} +c_3$, where
$c_2 ,c_3 \in \CC$ are independent of $\Lambda$ and
$\Omega^{\tw}$ is the Casimir operator of $\hat{\fg}^{\tw}$.  But
$\Omega^{\tw} v_{\Lambda}=(\Lambda |\Lambda
+2\hat{\rho}^{\tw})v_{\Lambda}$ (see \cite{K3}) and $Dv_{\Lambda}=0$, hence,
comparing with (\ref{eq:3.12}) we obtain for any
  $\Lambda \in \fh^{\sigma *}$:
\begin{displaymath}
\frac{1}{2(k+h\spcheck)} \left((\Lambda |\Lambda)-2(\Lambda
  |\gamma') \right)+c_1=c_2 \left( (\Lambda |\Lambda)
  +2(\hat{\rho}^{\tw}|\Lambda)\right) +c_3.
\end{displaymath}
Comparing quadratic
terms in $\Lambda$ we obtain $c_2=(2k+2h\spcheck)^{-1}$.
Comparng linear terms in $\Lambda$, we get $\hat{\rho}^{\tw}|_{\fh^{\sigma}}=
-\gamma'$.
\end{proof}

Recall that the conformal weight $1$ formal distributions of the
vertex algebra $W_k (\fg ,x)$ are \cite{KW}:
\begin{displaymath}
  J^{\{ v \}} (z) = J^{(v)} (z) -\frac{1}{2}\sum_{i,j \in S_{1/2}}
     (-1)^{p(u_i)} c_{ij}(v) : \Phi_i (z) \Phi^j (z) : \, (v\in\fg_0^f) .
\end{displaymath}
Hence, by Equation~(\ref{eq:1.14}), the corresponding twisted formal
distributions of $W_k (\fg ,\sigma, x)$ can be explicitly
expressed via twisted currents and twisted ghosts.  In the sequel
we shall need the following formula, in the case when $a \in
\fh^{\sigma f}$:
\begin{equation}
  \label{eq:3.13}
   J^{\{ a \}\tw}_0 =a-\sum_{i \in S_+}(-1)^{p(u_i)}
       s_i c_{ii}(a) +\frac{1}{2} \sum_{i \in S_{1/2}}
       (-1)^{p(u_i)}s_ic_{ii}(a) +\ann \, ,
\end{equation}
where $\ann$ denotes an operator which annihilates any vector of
the form $v \otimes \vac \in M \otimes F^{\tw}$. Formula (\ref{eq:3.13})
implies the following corollary.

\begin{corollary}
  \label{cor:3.3}
Under the conditions of Corollary~\ref{cor:3.1}, the eigenvalue
of $J^{\{ H \},\tw}_0$ $(H \in \fh^{\sigma f})$ on the vector $v
\otimes \vac$ is equal to
\begin{displaymath}
  \Lambda (H)+\frac{1}{2}\sum_{\alpha\in S_{1/2}}
  (-1)^{p(\alpha)} s_{\alpha} \alpha
  (H)-\sum_{\alpha \in S_+}
  (-1)^{p(\alpha)} s_{\alpha} \alpha (H)\, .
\end{displaymath}

\end{corollary}

As in \cite{KRW}, define the Euler-Poincar\'e character of
$H^{\tw}(M)$ by the following formula, where $h \in \fh^{\sigma
  f}$ and $\tau \in \CC$, $\Im \tau >0$:
\begin{displaymath}
  \ch_{H^{\tw}(M)} (\tau ,h) =\sum_{j \in \ZZ}(-1)^j
     \tr_{H_j(M)} e^{2\pi i \tau L^{\tw}_0} e^{2\pi i J^{\{ h
         \}}_0}\, .
\end{displaymath}
The same argument as in \cite{KRW} gives an explicit formula in
terms of the character
\begin{displaymath}
     \ch_M (\tau ,z)=\tr_M
     e^{2\pi i (z+\tau L^{\fg ,\tw}_0)}\, ,
     z \in \fh^{\sigma} \, ,
\end{displaymath}
of the $\hat{\fg}^{\tw}$-module $M$:
\begin{eqnarray}
  \label{eq:3.14}
\lefteqn{ \hspace{-3in} \ch_{H^{\tw}(M)}(\tau ,h) = e^{2\pi i \tau (s_{\ne}+s_{\ch})}}\\
\nonumber
\times \Biggl( \ch_M  \prod_{\substack{\alpha \in \hat{\Delta}_+ \\
      \alpha (x) \neq 0 ,-1/2}}
 (1-\tilde{p} (\alpha)
      e^{-\alpha})^{\tilde{p} (\alpha) \mult \alpha}\Biggr)
 (\tau ,-\tau x+h)
%
\end{eqnarray}
Here and further, in order to simplify notation, we let
$\tilde{p} (\alpha)=(-1)^{p(\alpha)}$, and for
$\alpha \in \hat{\fg}^*$,
we define
$\alpha (\tau , z) =2\pi i \alpha ( z-\tau D)$.

The conditions of non-vanishing of $\ch_{H^{\tw}(M)}$ are similar
to those in the non-twisted case \cite{KRW}.  Namely, the same
argument as in \cite{KRW}, Theorem~3.2, gives the following
result.

\begin{proposition}
  \label{prop:3.3}
Let $M$ be a restricted $\hat{\fg}^{\tw}$-module of level $k \neq
-h\spcheck$ and assume that $\ch_M (\tau ,h)$ extends to a
meromorphic function on the upper half space $\Im \tau >0$, $h
\in \fh^{\sigma}$, with at most simple poles at the hyperplanes
$\alpha =0$, where $\alpha$ are real even roots.  Then $
\ch_{H^{\tw}(M)} (\tau, h)$ is not identically zero iff the
$\hat{\fg}^{\tw}$-module $M$ is not locally nilpotent with
respect to all root spaces $\hat{\fg}_{-\alpha}$, such that
$\alpha$ are positive even real roots satisfying the following
three properties:
\begin{displaymath}
\hbox{(i)~~}
\alpha (D+x)=0 \, , \quad \hbox{(ii)~~} \alpha
|_{(\fh^{\sigma})^f}=0\, , \quad
\hbox{(iii)~~} \alpha (x) \neq 0 \, , \, -1/2\, .
\end{displaymath}

\end{proposition}
We shall use formula (\ref{eq:3.14}) and \cite{KW2,KW3} to
compute the characters of $W_k (\fg ,\sigma, x)$-modules in a
subsequent paper \cite{KW4} (cf. \cite{FKW,KRW}).

\begin{remark}
  \label{rem:3.1}
A slightly more explicit form of (\ref{eq:3.14})
is as follows:
\begin{gather*}
\ch_{H^{\tw}(M)} (\tau ,h) = e^{2\pi i \tau (s_{\ne}+s_{\ch})}\\
   \times   \Bigl( ch_M \prod_{\substack{\alpha \in S'\\ \alpha(x)>0}}
        \prod^{\infty}_{n=1}  \left((1-\tilde{p}(\alpha)
              e^{-(n-s_{\alpha})\delta +\alpha})
              (1-\tilde{p}(\alpha) e^{-(n-1+s_{\alpha})
                \delta-\alpha})\right)^{ \tilde{p}(\alpha)\mult\alpha}\\
    \times \prod_{\substack{\alpha \in S'\\
           \alpha (x) =1/2}}
       \prod^{\infty}_{n=1}
          (1-\tilde{p}(\alpha) e^{-(n-s_{\alpha})\delta
            +\alpha})^{-\tilde{p}(\alpha)\mult\alpha} \Bigr)   (\tau ,-\tau x+h) \, .
\end{gather*}
%
\end{remark}

Let $a \in (\fg^{\bmu}_{-j})^f$, $j \geq 0$, and let $J^{\{ a \} ,\tw}(z)$
 be the corresponding formal distribution of $W_k (\fg ,\sigma
 ,x)$ (see Theorem~\ref{th:2.1}).  As in the non-twisted case
 \cite{KW}, its conformal weight with respect to $L^{\tw}(z)$
 equals $\Delta_a =j+1$.  We therefore write
 \begin{equation}
   \label{eq:3.15}
   J^{\{ a \} ,\tw} (z) =\sum_{n \in \bmu -\Delta_a}
      J^{\{ a \} ,\tw} z^{-n-\Delta_a}\, .
 \end{equation}
Recall the isomorphism as $\fg^f_0$-modules $\fg^f \cong \fg_0
+\fg_{1/2}$ given by \cite{KW}, (\ref{eq:1.12}).  We shall identify
$\fg_0$ (and its subspace $\fh^{\sigma}$) with a $\sigma$-invariant 
subspace of $\fg^f$, using this
isomorphism.  A $W_k (\fg ,\sigma ,x)$-module $M$ is called a
\emph{highest weight module} with highest weight $\lambda \in
(\fh^{\sigma})^*$ if there exists a non-zero vector $v_{\lambda
} \in M$ such that:
\begin{equation}
  \label{eq:3.16}
  \hbox{\quad\quad polynomials in the operators }J^{\{ a \} ,\tw}_n
  \hbox{ applied to }v_{\lambda} \hbox{ span }M \, ,
\end{equation}
\vspace{-4ex}
\begin{eqnarray}
  \label{eq:3.17}
 && J^{\{ a\} ,\tw}_0 v_{\lambda} =
     \lambda (a) v_{\lambda} \hbox{ if } a \in \fh^{\sigma}\, ,
     \\[1ex]
  \label{eq:3.18}
 && J^{\{ a \} ,\tw}_m v_{\lambda} =
  0 \hbox{ if } m>0 \hbox{ or } m=0 \hbox{ and }
     a \in \fn_0 (\sigma)_+ \, .
\end{eqnarray}
The Verma module is defined in the same way as in \cite{KW}, and we
have the following twisted analogue of Theorem~6.3 from
\cite{KW}.

\begin{theorem}
  \label{th:3.1}
If $P$ is a Verma module over the Lie superalgebra
$\hat{\fg}^{\tw}$, then $H(P) =H_0 (P)$, and it is a Verma module
over $W_k (\fg ,\sigma ,x)$.
\end{theorem}

\section{Modules over $W_k (\fg ,\sigma ,\theta /2)$, the free
  field realizations and determinant formulas}
\label{sec:4}

Of particular interest are the vertex algebras $W_k (\fg ,\theta
/2)$ associated to a minimal gradation of $\fg$ \cite{KRW,KW}
(cf.~\cite{FL}).  In this case $\fg$ is one of the simple Lie
superalgebras $\sl (m|n)/\delta_{m,n} \CC I$, $osp (m|n)$ (=$spo(n|m)$),
$D(2 , 1\, ;\,  a)$, $F(4)$, $G(3)$ or one of the five exceptional Lie
algebras, $\theta$ is the highest root of one of the simple
components of the even part of $\fg$, the bilinear form $(\, . \,
|\, . \, )$ is normalized by the condition $(\theta |\theta )=2$,
and $x=\theta /2$.  The corresponding $\frac{1}{2} \ZZ $-gradation
(\ref{eq:2.1}) looks as follows:
\begin{equation}
  \label{eq:4.1}
  \fg=  \CC f +\fg_{-1/2}+\fg_0 +\fg_{1/2}+\CC e \, ,
\end{equation}
where $\{ e,x,f \}$ form an $\sl_2$ triple , and
\begin{equation*}
\fg^f_0 = \{ a \in \fg_0 | (x|a) =0 \} \, , \quad
\fg^f = \CC f + \fg_{-1/2}+\fg^f_0\, .
\end{equation*}
Then
\begin{displaymath}
  \fg (\sigma)=\CC f + \fg^{-\sigma}_{-1/2}+\fg^{\sigma}_0
     +\fg^{-\sigma}_{1/2} +\CC e
\end{displaymath}
is a minimal gradation of $\fg (\sigma)$.  Since
$\fg^{\sigma}_0=(\fg^{\sigma}_0)^f + \CC x$, it follows that
there exists an element $h_0 \in \fh^{\sigma f}$ of the Lie
superalgebra $\fg (\sigma)$ such that the eigenvalues of $\ad h_0$
are real, $h_0$ is a regular element of $\fg^{\sigma}_0$, and the
$0$\st{th} eigenspace of $\ad h_0$ on $\fg^{-\sigma}_{1/2}$
(resp. $\fg^{-\sigma}_{-1/2}$) is $\CC e_{\theta /2}$ (resp. $\CC
e_{-\theta /2}$) if $e_{\theta /2}$ is a root vector of $\fg
(\sigma)$. 
(Here $\theta /2$ stands for the restriction of $\theta /2$ to $\fh^{\sigma}$.) 
 Letting $\fn_+ (\sigma)$ (resp. $\fn_- (\sigma)$) be
the span of all eigenvectors of $\ad h_0$ with positive
(resp. negative) eigenvalues and the vectors $f=e_{-\theta}$ and
$e_{-\theta/2}$ (resp. $e=e_{\theta}$ and $e=e_{\theta /2}$), we
obtain the decomposition (\ref{eq:2.4}), satisfying the properties
(i)---(iv).  Note also that in the decomposition (\ref{eq:2.5}),
$\fh_{1/2}(\sigma)$ (resp. $\fn_{1/2}(\sigma)'$) is the span of
all eigenvectors of $\ad h_0$ with positive (resp. negative)
eigenvalues, and $\fg_{1/2}(\sigma)_0 =\CC e_{\theta /2} \in \fg
(\sigma)$.  
Thus, $\epsilon (\sigma) \neq 0$ iff $\theta /2$ is a
root of $\fg$  with respect to $\fh^{\sigma}$ and $\sigma (e_{\theta /2})=-e_{\theta /2}$.

\begin{example}
  \label{ex:4.1}
For the minimal gradation the numbers $s_{\alpha} (\alpha \in
\Delta \subset \fh^*)$ are as follows (cf.~(\ref{eq:3.2})):

\alphaparenlist

\begin{enumerate}
\item 
If $\sigma =1$, then $s_{\alpha} =0$ (resp.~$1$) for $\alpha \in
\Delta_+$ (resp.~$-\alpha \in \Delta_+$).

\item 
If $\sigma |_{\fg_j} =(-1)^{2j}$, then $s_{\alpha}=0$ (resp.~$1$)
if $\alpha (x) =0$ and $\alpha (h_0) >0$ (resp.~$\alpha (h_0)
<0$), $s_{\alpha} =-1/2$ (resp.~$\frac{1}{2}$) if $\alpha (x)
=\frac{1}{2}$ and $\alpha (h_0) >0$ (resp.~$\leq 0$),
$s_{\theta} =0$ and $s_{\alpha} +s_{-\alpha}=1$, $\alpha \in \Delta$.
\end{enumerate}

\end{example}

Recall that the (Virasoro) central charge of $W_k(\fg,\theta/2)$ is \cite{KW}:
\begin{displaymath}
c(k)=\frac{k \sdim \fg}{k+h\spcheck}-6k+h\spcheck-4\, ,
\end{displaymath}
and it is, of course, the same for the twisted vertex algebras $W_k(\fg,\sigma,\theta/2)$.

Introduce the following vectors in $\fh^{\sigma *}$:
\begin{displaymath}
  \gamma'=\frac{1}{2}\sum_{\alpha \in S'} \tilde{p}(\alpha)
      s_{\alpha} \alpha \, , \quad
  \gamma_{1/2} =\frac{1}{2}\sum_{\alpha \in S_{1/2}}
      \tilde{p} (\alpha)s_{\alpha} \alpha \, .
\end{displaymath}

Corollaries \ref{cor:3.1} and \ref{cor:3.3}
give the following result, which will be used in the calculation
of the determinant formula.

\begin{proposition}
  \label{prop:4.1}
Let $M$ be a restricted $\hat{\fg}^{\tw}$-module of level~$k$.
Let $v \in M$ be a singular vector of $M$ with weight $\Lambda
\in \fh^{\sigma *}$. Then  we have in the
case of $W_k (\fg ,\sigma\, , \,  \theta /2)$:

\alphaparenlist
\begin{enumerate}
\item 
The eigenvalue of $L^{\tw}_0$ on $v \otimes \vac$ is equal to
\begin{displaymath}
  h = \frac{1}{2(k+h\spcheck)} ((\Lambda | \Lambda)
  -2(\Lambda|\gamma'))
%
  -\Lambda(x) + s_{\fg} + s_{\gh}\, ,
\end{displaymath}
where
\begin{displaymath}
  s_{\fg}=-\frac{k}{4(k+h\spcheck)} \sum_{\alpha \in S'}
     \tilde{p}(\alpha) s_{\alpha} (s_{\alpha}-1)\, , \quad
     s_{\gh}=\frac{1}{4}
     \sum_{\alpha \in S_{1/2}} \tilde{p}(\alpha) s^2_{\alpha}\, .
\end{displaymath}
\item 
The eigenvalue of $J^{\{ H \} ,\tw}$ for $H \in \fh^{\sigma f}$ on $v
\otimes \vac$ is equal to
\begin{displaymath}
 (\Lambda - \gamma_{1/2}) (H) \, .
\end{displaymath}
\end{enumerate}
\end{proposition}

\begin{proof}
  Letting $s_{\gh}:=s_{\ch}+s_{\ne}$, we have (see (\ref{eq:3.10})
  and (\ref{eq:3.11})):  $s_{\gh}=\frac{1}{8}\epsilon (\sigma)
  +\frac{1}{4} \sum_{\alpha \in S_{1/2}}\tilde{p}
  (\alpha)(s^2_{\alpha}+s_{\alpha})$.  Since $\alpha \in S_{1/2}$ iff $\theta -\alpha \in S_{1/2}$, we
obtain that
\begin{equation}
\label{eq:4.2}
\sum_{\alpha \in S_{1/2}} \tilde{p}(\alpha)
s_{\alpha}=-\frac{1}{2}\epsilon (\sigma)\, .
\end{equation}
It is because
$s_{\alpha}+s_{\theta-\alpha}=\delta_{\alpha ,\theta/2}$, which
holds due to (\ref{eq:3.6}). This proves the formula for
$s_{\gh}$.  The rest is straightforward.

\end{proof}

\begin{proposition}
  \label{prop:4.2}
For the $\frac{1}{2}\ZZ$-gradation of $\fg$ defined by $\ad x$
one has:

\alphaparenlist
\begin{enumerate}
\item 
$2\gamma' (x) =1-h\spcheck -\frac{1}{2}\epsilon (\sigma)$.

\item 
$\gamma' =2\gamma_{1/2} +\gamma'_0 - \frac{1}{2}(h\spcheck -1)\theta$, where
$\gamma'_0 =\frac{1}{2}\sum_{\substack{\alpha \in
    S'\\\alpha(x)=0}} \tilde{p} (\alpha) s_{\alpha}\alpha$.

\item 
$\gamma^{\natural}_{1/2}=\frac{1}{2}(\gamma^{\prime \natural} -\gamma^{\prime \natural}_0)$.
\end{enumerate}

\end{proposition}

\begin{proof}
  We have:  $2\gamma' (x) = \frac{1}{2}\sum_{\alpha \in S_{1/2}}
  \tilde{p} (\alpha)s_{\alpha} -\frac{1}{2}\sum_{\alpha \in S_{-1/2}}
  \tilde{p} (\alpha) s_{\alpha}-1=\sum_{\alpha \in S_{1/2}}
    \tilde{p}(\alpha) s_{\alpha} -\frac{1}{2} \sdim \fg_{1/2} -1$.
 Since $\sdim \fg_{1/2}=2h\spcheck
-4$ (see \cite{KW}, (5.6)), formula~(\ref{eq:4.2}) completes the proof of~(a).

Similar calculations establish~(b), and (c) is immediate by~(b).
\end{proof}

In \cite{KW}, Theorem 5.2, we gave a realization of the vertex algebra\break $W_k
(\fg ,\theta/2)$ as a subalgebra of $V_{\alpha_k} (\fg_0)
\otimes F (A_{\ne})$, where $\fg_0$ is the $0$\st{th} grading component
in (\ref{eq:2.1}) and $\alpha_k$ is the ``shifted'' $2$-cocycle:
$\alpha_k(at^m,bt^n)=((k+h\spcheck)(a|b)-\frac{1}{2}\kappa_{\fg_0}(a,b))m\delta_{m,-n}$,
where $\kappa_{\fg_0}$ is the Killing form on $\fg_0$.  The twisted
version of this result
is derived from \cite{KW}, Theorem~5.2, by
making use of (\ref{eq:1.14}), Theorem~\ref{th:2.1}, and the
following identity for formal distributions $a,b,c$ such that
$[a_{\lambda}b]= \langle a,b \rangle \in \CC,
  [a_{\lambda} c]=\langle a ,c \rangle \in \CC,
  [b_{\lambda}c]=\langle b,c \rangle \in \CC$:
\begin{eqnarray}
\label{eq:4.3}
:abc :^{\tw}(z) &=& :a^{\tw} (z) b^{\tw}(z) c^{\tw}(z):
   - z^{-1} \big( s_b \langle b,c \rangle a^{\tw} (z) +\\
\nonumber
 && s_{a} \langle a,b \rangle c^{\tw}(z) + s_a (-1)^{p(a)p(b)}
   \langle a,c \rangle b^{\tw} (z)\big) \, .
\end{eqnarray}
As in \cite{KW}, we keep the notation $J^{\{a\},\tw}$ if $a\in\fg_0^f$, but let $G^{\{v\},\tw}
=J^{\{v\},\tw}$ if $v\in \fg_{-1/2}$. Due to Theorem~\ref{th:2.1}, the formal distributions
$J^{\{a\},\tw}$, $G^{\{v\},\tw}$ and $L^{\tw}$ strongly and freely generate the twisted vertex 
algebra $W_k(\fg,\sigma,\theta/2)$.

\begin{theorem}
  \label{th:4.1}
The following formulas define an injective vertex algebra homomorphism of  $W_k(\fg ,\sigma
,\theta/2)$ to $V_{\alpha_k} (\fg_0 ,\sigma) \otimes F
(A_{\ne},\sigma)$:
\begin{multline*}
 J^{\{ a \},\tw}(z) \mapsto a^{\tw}(z) + \frac{(-1)^{p(a)}}{2}
    \sum_{\alpha \in S_{1/2}} :\Phi^{\alpha,\tw}(z)
      \Phi^{\tw}_{[u_{\alpha},a]} (z):\\
-\frac{(-1)^{p(a)}}{2}\sum_{\alpha \in S_{1/2}} s_{\Phi^{\alpha}}
   \langle \Phi^{\alpha} ,\Phi_{[u_{\alpha} ,a]} \rangle_{\ne}z^{-1}
   (a \in \fg^f_0)\, , \\
G^{\{ v \}, \tw} (z) \mapsto  \sum_{\alpha \in S_{1/2}}:[v ,u_{\alpha}]
   (z) \Phi^{\alpha,\tw} (z) :-(k+1) \sum_{\alpha \in S_{1/2}}
      (v |u_{\alpha}) \partial \Phi^{\alpha,\tw} (z)\\
-\frac{(-1)^{p(v)}}{3} \sum_{\alpha,\beta \in S_{1/2}}:\Phi^{\alpha,\tw}
   (z) \Phi^{\beta,\tw}(z) \Phi^{\tw}_{[u_{\beta},[u_{\alpha},v]]} (z) :\\
+\frac{(-1)^{p(v)}}{3} \sum_{\alpha,\beta \in S_{1/2}}
   \left( s_{\Phi^{\beta}} \langle \Phi^{\beta}, \Phi_{[u_{\beta},[u_{\alpha},v]]}
   \rangle_{\ne} \Phi^{\alpha,\tw}(z) \right. \\
+ (-1)^{p(\alpha)p(\beta)} s_{\Phi^{\alpha}} \langle  \Phi^{\alpha},
\Phi_{[u_{\beta},[u_{\alpha},v]]}
   \rangle_{\ne} \Phi^{\beta,\tw}(z) \\
  \left. + s_{\Phi^{\alpha}} \langle \Phi^{\alpha},\Phi^{\beta} \rangle_{\ne}
 \Phi^{\tw}_{[u_{\beta},[u_{\alpha},v]]} (z)\right) z^{-1} \,\, (v \in
\fg_{-1/2})\, ,
\end{multline*}

\begin{multline*}
L^{\tw}(z) \mapsto \frac{1}{2(k+h\spcheck)} 
\sum_{\alpha \in S_0}
(-1)^{p(\alpha)} : u^{\tw}_{\alpha} (z) u^{\alpha,\tw} (z) :+\frac{k+1}{k+h\spcheck} \partial x(z)\\
+\frac{1}{2} \sum_{\alpha \in S_{1/2}} (-1)^{p(\alpha)}:\Phi^{\tw}_{\alpha}
   (z) \partial \Phi^{\alpha,\tw}(z) : 
-\frac{1}{2(k+h\spcheck)} \sum_{\alpha \in S_0'}(-1)^{p(\alpha)}
   s_{\alpha} \alpha (z)z^{-1}\\
+ \Big(\frac{1}{4(k+h\spcheck)} \sum_{\alpha \in S_0}(-1)^{p(\alpha)}
    \binom{s_{\alpha}}{2}
    \kappa_{\fg_0}(u_{\alpha},u^{\alpha})
    -\frac{1}{2}\sum_{\alpha \in S_+\cup S_0}
      (-1)^{p(\alpha)} \binom{s_{\alpha}}{2} \Big) z^{-2} \, .
\end{multline*}
(For $\fg^f_0$ simple, $\kappa_{\fg_0}(u_{\alpha},u^{\alpha})=2 h\spcheck_0$,where 
$h\spcheck_0$ is the dual Coxeter number of $\fg^f_0 $ with respect to $(\, . \, | \, . \, )$.)

\end{theorem}

In the case of $W_k (\fg , \sigma ,\theta /2)$,
Proposition~\ref{prop:3.1} gives the following result.

\begin{proposition}
  \label{prop:4.3}
Let $M$ be a $\hat{\fg}^{\tw}$-module satisfying the conditions
of Proposition~\ref{prop:3.1}.  Then
the Euler-Poincar\'e
character of the $W_k (\fg ,\sigma ,\theta /2)$-module $H^{\tw}
(M)$ is not identically zero iff $e_{\theta}t^{-1}$ is
not locally nilpotent on $M$.
\end{proposition}

Now we turn to the determinant formula for the Verma modules over
$W_k (\fg ,\sigma ,\theta/2)$.  To simplify notation, we let
$\fg^{\natural} =\fg^f_0$ (resp.~
$\fh^{\natural}=(\fh^{\sigma})^f$), the centralizer of $f$ in $\fg_0$
(resp. in $\fh^{\sigma}$).  Let $\lambda \mapsto \lambda^{\natural}$ denote
the restriction map $\fh^{\sigma}\to \fhs$. Let $S_0$
(resp.~$S_{-1/2}$) $=\{ \alpha \in S' |\, \alpha (x) =0$
(resp.~$\alpha (x) =-1/2) \}$, and let
$\Delta^{\natural}_{W,\sigma} =\{ \alpha^{\natural} |\, \alpha
\in S_0 \cup S_{-1/2} \} \subset \fh^{\natural *}$, the multiplicity of
$\alpha^{\natural}$ being the multiplicity of $\alpha \in S'$.
Note that $\Delta^{\natural}_{W,\sigma}$ may contain $0$ (this
happens iff $\theta /2 \in \Delta^{\sigma}$).

 Define the set of roots $\Delta_{W,\sigma}$ of
$W_k(\fg,\sigma ,\theta/2)$ as a subset of the dual of its Cartan
algebra
\begin{displaymath}
  \fh_{W,\sigma} =\fhs \bigoplus \CC L^{\tw}_0 \, ,
\end{displaymath}
defined as follows.  We embed $
\fh^{\natural *}$ in $\fh^*_{W,\sigma}$ by letting $\alpha \in
\fh^{\natural *}$ to be
zero on $L^{\tw}_0$, and define $\delta' \in \fh^*_{W,\sigma}$ by
\begin{displaymath}
  \delta' |_{\fh^{\natural}}=0 \, , \quad \delta' (L^{\tw}_0)=-1\, .
\end{displaymath}
Then $\Delta_{W,\sigma} =\Delta^{\re}_{W,\sigma} \cup
\Delta^{\im}_{W,\sigma}$, where
\begin{eqnarray*}
  \Delta^{\re}_{W,\sigma} &=& \{ (n+s_{\alpha}+\alpha (x))\delta'
     +\alpha \, | \,\, \alpha \in \Delta^{\natural}_{W,\sigma} \, ,
     \quad n \in \ZZ \} \, , \\
   \Delta^{\im}_{W,\sigma} &=& \{ n\delta' | \,\, n \in E_0 \, ,
   \quad n \neq 0 \} \, ,
\end{eqnarray*}
where the multiplicity of a root $(n+s_{\alpha}+\alpha (x))\delta'
+\alpha$ is equal to the multiplicity of $\alpha \in
\Delta^{\natural}_{W,\sigma}$ with given $s_{\alpha}$, and the multiplicity of $n\delta'$ is
equal to the multiplicity of the root $n\delta$ of $\hat{\fg}^{\tw}$.  Note
that $0$ is a root in $\Delta^{\re}_{W,\sigma}$ of multiplicity
$\epsilon (\sigma) (\leq 1)$.

We denote by $\Delta^+_{W,\sigma}$ the subset of positive roots,
which consists of the subset
$\Delta^{\im,+}_{W,\sigma}$ of
elements of $\Delta^{\im}_{W,\sigma}$ for
which $n >0$, and the subset
$\Delta^{re,+}_{W,\sigma}$ of elements of
$\Delta^{re}_{W,\sigma}$ for which
$n \in \ZZ_+$.

Define the corresponding partition function $P_{W,\sigma}(\eta)$
on $\fh^*_{W,\sigma}$ as the number of ways $\eta$ can be
represented in the form (counting root multiplicities):
\begin{displaymath}
  \eta = \sum_{\alpha \in \Delta^+_{W,\sigma}} k_{\alpha}\alpha \, ,
  \hbox{ where } k_{\alpha}\in \ZZ_+ \hbox{ and }k_{\alpha}\leq 1
  \hbox{ if }\alpha \hbox{ is odd.}
\end{displaymath}

\begin{remark}\hspace{-1.5ex} 
 \label{re:4.1}
Denote by $P'_{W,\sigma} (\eta)$ the partition 
function for the set $\Delta^{+}_{W,\sigma} \backslash \{ 0 \}$. Of course, $P'_{W,\sigma} (\eta) =P_{W,\sigma}(\eta)$ if $\epsilon (\sigma)=0$, but $P'_{W,\sigma} (\eta) =\tfrac{1}{2}P_{W,\sigma} (\eta)$ if $\epsilon (\sigma)=1$ and $\eta \neq 0$.
\end{remark}

The definition (\ref{eq:3.16})---(\ref{eq:3.18}) of  a highest
weight module $M$ over the vertex algebra $W_k(\fg ,\sigma ,\theta /2)$ 
can be made
a bit more explicit:  the highest weight $\lambda$ is an element
of $\fh^*_W$, and condition (\ref{eq:3.17}) can be replaced by
\begin{equation}
  \label{eq:4.4}
  J^{\{ H \}}_0 v_{\lambda}= \lambda^{\natural} (H) v_{\lambda}\, ,
    H \in \fhs \, , \hbox{ and } L^{\tw}_0
     v_{\lambda} =hv_{\lambda}\, ,
\end{equation}
where $\lambda^{\natural}$ denotes the restriction of $\lambda$ to
$\fhs$ and $h$ is the minimal eigenvalue  of $L^{\tw}_0$ on $M$.
We have the weight space decomposition of $M$:
\begin{eqnarray*}
  M=\bigoplus_{\mu \in \fh^*_{W,\sigma}} M_{\mu}\, , \,
  M_{\mu} =\{ v \in M |J^{\{ H \}}_0v
  =\mu^{\natural} (H) v \, , \, H\in \fhs \, , \, L^{\tw}_0
  v = \mu (L^{\tw}_0) v \}\, .
\end{eqnarray*}
The Verma module $M(\lambda)$ over $W_k (\fg ,\sigma,x)$ is a
highest weight module for which
\begin{displaymath}
  \dim M (\lambda)_{\mu} =P_{W,\sigma}(\lambda -\mu)\, .
\end{displaymath}

In the case when $\epsilon (\sigma)(=\dim \fg_{1/2}(\sigma)_0)=1$
choose a non-zero vector $e' \in \fg_{1/2}(\sigma)_0$ and let
$f'=[f,e']$. Rescaling $e'$, if necessary, we may assume that $[f',f']=f$.
The vector $f'$ is a
weight vector for $\fhs$ in $\fg_{-1/2}$ with weight zero.  
Due to Theorem~\ref{th:2.1}, we have the
corresponding formal distribution in $W_k (\fg ,\sigma ,\theta/2)$:
\begin{equation}
  \label{eq:4.5}
  G(z) :=(-k-h\spcheck)^{-1/2} G^{\{f'\} ,\tw}(z) = \sum_{n \in \ZZ}
      G_n z^{-n-3/2} \, .
\end{equation}
We have the following description of the highest weight subspace
of $M(\lambda)$:
\begin{eqnarray*}
  M(\lambda)_{\lambda} =\left\{
    \begin{array}{l}
      \CC v_{\lambda} \hbox{ if }\epsilon (\sigma)=0 \, ,\\
      \CC v_{\lambda} +\CC G_0 v_{\lambda} \hbox{ if }
         \epsilon (\sigma) =1\, .
    \end{array}\right.
\end{eqnarray*}

We shall need an explicit formula for the eigenvalue of
$[G_0,G_0]$ on $v_{\lambda}\in M(\lambda)$, which we shall denote
by $\varphi_0 (k,h,\lambda^{\natural})$.  In order to compute the
function $\varphi_0$, recall that Theorem~5.1(e) from \cite{KW}
provided an explicit expression for $[{G^{\{ u\}}}_{\lambda} G^{\{v \}}]$, 
$u,v \in \fg_{-1/2}$, in $W_k (\fg ,\theta /2)$.
Unfortunately, the coefficient of $\frac{\lambda^2}{6}$ in this
expression is correct only when $\fg^{\natural} =\fg^f_0$ is
simple.  Here is a correct expression for this coefficient, which
we shall denote by $\gamma_{k}$:
\begin{eqnarray*}
  \gamma_k (u,v) =-(k+h\spcheck) g(u,v)c(k) 
     + g (u,v)\sum_{\alpha \in S^{\natural}} \beta_k 
     (u^{\alpha} ,u_{\alpha})\\
     + 2\sum_{j \in S_{1/2}}\beta_k ([u,u^j]^{\natural} \, , \, 
     [u_j,v]^{\natural})\, ,
\end{eqnarray*}
where $\beta_k (a,b)=(k+\frac{1}{2}h\spcheck) (a|b)-\frac{1}{4}
\kappa_{\fg_0} (a,b),a,b \in \fg_0$, $g(u,v) \in \CC$ is defined by
$[u,v]=g(u,v)f$, $a^{\natural}$ stands for the orthogonal
projection of $a \in \fg_0$ on $\fg^{\natural}$, and
$S^{\natural}$ indexes a basis of $\fg^{\natural}$.  If
$\fg^{\natural}$ is simple or, more generally, if $\kappa_{\fg_0}
(a,b)=2h\spcheck_0 (a|b),a,b \in \fg^{\natural}$, we have a much
simpler formula:
\begin{displaymath}
  \gamma_k (u,v)=-g(u,v)\Big(\big( k+h\spcheck \big)c(k)- 
  \big( k+\frac{1}{2}(h\spcheck -h\spcheck_0)\big)
  ( \sdim \fg_{0} +\sdim \fg_{1/2})\Big) \, .
\end{displaymath}

In the case when $u=v=f'$, so that $g(u,v)=1$, we obtain from
\cite{KW}, Theorem~5.1(e):
\begin{eqnarray*}
\lefteqn{\hspace{-4in}  [{G^{\{ f'\}}}_{\lambda} G^{\{ f'\}}] = - (k+h\spcheck)L} \\
     +\frac{1}{2}\Big( \sum_i :J^{\{ h^i \}}J^{\{h_i\}}:
     +\sum_{\alpha \in S ''_0} :J^{\{ u^{\alpha}\}}J^{\{u_{\alpha}\}}:)
     +\frac{\lambda^2}{6}\gamma_k (f',f')\, .
\end{eqnarray*}
Here $\{h_i \}$ and $\{ h^i\}$ are dual bases of $\fh^{\natural}$,
and $S''_0$ is a basis of the kernel of the map  $\ad f'
:\sum_{\alpha \in S'_0} \CC u_{\alpha} \to \fg_{-1/2}$.

This formula is used to obtain:
\begin{eqnarray}
  \label{eq:4.6}
\lefteqn{\hspace{-4in}   \varphi_0(k,h,\lambda^{\natural}) =h-\frac{1}{2(k+h\spcheck)}
    \Big( |\lambda^{\natural}
    +\gamma^{\natural}_{1/2}-\gamma^{\prime\natural}|^2}\\
\nonumber
 \lefteqn{\hspace{-4in}   - |\gamma^{\natural}_{1/2} -\gamma^{\prime\natural}|^2
    -\sum_{\alpha \in S''_0} (-1)^{p(\alpha)}
    \binom{s_{\alpha}}{2}
 \beta_k (u_{\alpha},u^{\alpha})\Big)}\\
\nonumber
   -\frac{1}{24 (k+h\spcheck)} \Big(\sum_i \beta_k (h_i,h^i)
   +\sum_{\alpha \in S''_0} (-1)^{p(\alpha)} \beta_k
   (u_{\alpha}, u^{\alpha})\Big) -\frac{c(k)}{24}\, .
\end{eqnarray}
Note that in the case when $\kappa_{\fg_0} (a,b)=2h\spcheck_0 (a|b)$,
this formula can be simplified, using
$  \beta_k (u_{\alpha} ,u^{\alpha})=\beta_k(h_i,h^i)
    = k +\frac{1}{2}(h\spcheck -h\spcheck_0)$.  Then
    (\ref{eq:4.6}) becomes:
    \begin{gather}
      \label{eq:4.7}
      \varphi_0 (k,h,\lambda^{\natural}) 
        =  h-\frac{1}{2(k+h\spcheck)}\Big( |\lambda^{\natural}+
          \gamma^{\natural}_{1/2}-\gamma^{\prime\natural} |^2 \\
\nonumber
        -|\gamma^{\natural}_{1/2}-\gamma^{\prime\natural}|^2
        + \frac{h\spcheck +h\spcheck_0}{2}\sum_{\alpha \in S''_0}
        (-1)^{p(\alpha)} \binom{s_{\alpha}}{2}-\frac{1}{2}
        (k+\frac{1}{2})^2 \\
\nonumber
        +\frac{h\spcheck (h\spcheck -1)}{3}+\frac{1}{8}
        -\frac{h\spcheck -h\spcheck_0}{24}
        (\sdim \fg_0 +\sdim \fg_{1/2})\Big)\\
\nonumber
     \lefteqn{\hspace{-1.65in}   +\frac{1}{2}\sum_{\alpha \in S''_0} (-1)^{p(\alpha)}
        \binom{s_{\alpha}}{2}+\frac{h\spcheck}{8}\, .}
    \end{gather}

In order to define the contravariant bilinear form on a Verma
module $M(\lambda)$ over $W_k (\fg ,\sigma ,x)$, we use the
anti-involution $\omega$ of $\fg$ introduced in \cite{KW}; we
shall assume that it commutes with $\sigma$.  As in \cite{KW}, we
have the following anti-involution of the associative algebra
$\A$ generated by coefficients $J^{\{ a \} ,\tw}_n$ of formal
distributions $J^{\{ a \} ,\tw}(z)$, where $a \in \fg^{\natural} \oplus
\fg_{-1/2}$ (see~(\ref{eq:3.14})), and the $L^{\tw}_n$:
\begin{displaymath}
  \omega (L^{\tw}_n) =L^{\tw}_{-n} \, , \,
  \omega (J^{\{ a \} ,\tw}_n) =J^{\{ \omega (a)\},\tw}_{-n}\, .
\end{displaymath}
The contravariant bilinear form $B (\, . \, , \, .\, )$ on a
Verma module $M(\lambda)$ over $W_k (\fg ,\sigma ,x)$ with
highest weight vector $v_{\lambda}$ is defined in the usual
way:
\begin{displaymath}
  B(av_{\lambda}, bv_{\lambda}) =
    \langle v^*_{\lambda} \, , \, \omega (a)bv_{\lambda}
    \rangle \, , \quad a,b \in \A \, ,
\end{displaymath}
where $v^*_{\lambda}$ is the linear function on $M(\lambda)$,
equal to $1$ on $v_{\lambda}$ and $0$ on $G_0 v_{\lambda}$
and all weight spaces $M(\lambda)_{\mu}$, $\mu \neq \lambda$.
This is a symmetric bilinear  form, which is contravariant,
i.e.,~$B (au,v)=B(u, \omega (a)v)$, $u,v \in M(\lambda)$, $a \in \A$, and $B
 (v_{\lambda},v_{\lambda})=1$, and these properties
 determine  $B (\, . \, , \, .\, )$ uniquely.  Different weight
 spaces are orthogonal with respect to this form and its kernel
 is the maximal submodule of $M(\lambda)$.

Denote by $\det_{\eta} (k,h,\lambda^{\natural})$ the determinant of
the bilinear form  $B (\, . \, , \, .\, )$ restricted to the
weight space $M(\lambda)_{\lambda -\eta}$,  $\eta \in
\fh^*_{W,\sigma}$.  This is a function in $k$, $h$ and
$\lambda^{\natural}$ (see (\ref{eq:4.4})) and it depends on the choice
of a basis of $M(\lambda)_{\lambda -\eta}$ only up to
a constant factor.

Consider the map $\pi : \hat{\Delta}\to
\fh^*_{W,\sigma}$, defined by
\vspace{-1ex}
\begin{displaymath}
  \pi (\alpha + m\delta)=
    \alpha^{\natural} +(m +\alpha (x))\delta' \, , \,\,
    \pi (m\delta) =m\delta' \, .
\end{displaymath}
It is easy to see that, counting root multiplicities,
$\pi$ induces a bijective map:
\begin{displaymath}
  \pi (\hat{\Delta}^{\re}_{++} \cup \hat{\Delta}^{\im}_+)
 \overset{\sim}{\to}
       \hat{\Delta}^{+}_{W,\sigma} 
       \backslash \{0\}.
\end{displaymath}

Denote by $\mult_0 m\delta$ the multiplicity of
the root $m\delta$ in
$ \hat{\fg^{\natural}}^{\tw}$($ \subset \hat{\fg}^{\tw}$).

\begin{theorem}
  \label{th:4.2}\quad
Up to a non-zero constant factor, the determinant $\det_{\eta}$\break
$(k,h,\lambda^{\natural})$ is given by the following formula:
\begin{gather*}
  \varphi_0 (k,h,\lambda^{\natural})^{\epsilon (\sigma)P'_{W,\sigma}(\eta)}
  \prod_{\substack{m \in E^+_0\\ n \in \NN}} (k+h\spcheck)^{(\mult_0 m\delta )
P_{W,\sigma}(\eta-mn\delta')}\\[1ex]
\times \prod_{\substack{\alpha \in \hat{\Delta}^{\re}_{++}\\ n \in \NN}}
    \varphi_{\alpha ,n}(k,h,\lambda^{\natural})^{\tilde{p}(\alpha)^{n+1}
      P_{W,\sigma} (\eta -n \pi (\alpha) )}\, ,
\end{gather*}
where the factor $\varphi_0$ (occurring only when $\epsilon(\sigma)=1$)
is given by
(\ref{eq:4.6}), and all the remaining factors are as follows
($\alpha \in \Delta^{\sigma} , m \in E_0$):
\begin{alignat}{2}
  \label{eq:4.8}
\shoveleft{\varphi_{m\delta +\alpha ,n} \!
     =\!\!  m(k+h\spcheck)+
      (\lambda^{\natural} +\gamma_{1/2}^{\natural}-\gamma^{\prime \natural}|\, \alpha)-\!
      \frac{n}{2}|\alpha |^2
      \text{ if $\alpha (x)\!\!=0$,}}\\
  \label{eq:4.9}
 \lefteqn{ \hspace{-4.1in}\varphi_{m\delta +\alpha ,n} =}\hfill\\
     h-\frac{1}{k+h\spcheck}
        \Bigl(\bigl(\frac{n}{2}|\alpha |^2-(m+\frac{1}{2})(k+h\spcheck)
        -(\lambda^{\natural}+\gamma^{\natural}_{1/2}
        -\gamma^{\prime \natural}|\alpha)\bigr)^2 \notag \\
    + \frac{1}{2}|\lambda^{\natural}+\gamma^{\natural}_{1/2}
    -\gamma^{\prime \natural} |^2
  - \frac{1}{4}(k+1-\frac{1}{2}\epsilon(\sigma))^2
  -\frac{1}{2}|\gamma^{\prime\natural}|^2\Bigr)
  -s_{\fg}-s_{\gh} \notag\\
 \intertext{ if   $\alpha (x) =\frac{1}{2}$ \, ,}
  \label{eq:4.10}
\lefteqn{\hspace{-4in}\varphi_{m\delta +\theta ,n}= h-\frac{1}{4(k+h\spcheck)}
   \Bigl(\bigl(n-(m+1)(k+h\spcheck)\bigr)^2}\hfill\\
    +2|\lambda^{\natural} + \gamma^{\natural}_{1/2}-
      \gamma^{\prime \natural}|^2 - (k+1-\frac{1}{2}\epsilon(\sigma))^2 -
      2|\gamma^{\prime \natural}|^2\Bigr) -s_{\fg}-s_{\gh}\, . \notag
\end{alignat}
(Formulas for $s_{\fg}$ and $s_{\gh}$ are given in
Proposition~\ref{prop:4.1}(a).)

\end{theorem}

\begin{proof}
  The proof follows the traditional lines, as in \cite{KW}.
  First, let $M$ be a Verma module over $\hat{\fg}^{\tw}$ with
  highest weight $\hat{\Lambda}=\Lambda +kD$, where $\Lambda
  \in \fh^{\sigma *}$.  Then for each $\hat{\alpha} \in
  \hat{\Delta}_+$ and a positive integer $n$ such that
  \begin{equation}
    \label{eq:4.11}
    2(\hat{\Lambda} +\hat{\rho}^{\tw} |\hat{\alpha})
       =n(\hat{\alpha}|\hat{\alpha})
  \end{equation}
under certain conditions (stated in Lemma~7.1 of \cite{KW}),
$\hat{\Lambda}-n\hat{\alpha}$ is a singular weight, of
multiplicity at least $\mult \alpha$, of $M$.  This follows from
the determinant formula for $\hat{\fg}^{\tw}$ in \cite{K2} (as
corrected in Remark~7.1 of \cite{KW}).

Let ${h\spcheck}^\tw =(\hat{\rho}^{\tw}|\delta)$ be the dual
Coxeter number of $\hat{\fg}^{\tw}$.  We have:
\begin{equation}
  \label{eq:4.12}
  {h\spcheck}^{\tw}=h\spcheck\, .
\end{equation}
Indeed, the $L^{\tw}_n$ can be constructed for all $k \neq
-{h\spcheck}^{\tw}$ (see \cite{K3}, Exercise~12.20).  But the
central charge of $L^{\fg ,\tw}$ is independent of $\sigma$ and
has singularity only at $k=-h\spcheck$.  This implies
(\ref{eq:4.12}).  Hence for $\hat{\alpha}=\alpha +m\delta$, where
$\alpha \in \Delta^{\sigma}$, $m \in E_0$, (\ref{eq:4.11}) can be rewritten,
using also Corollary~\ref{cor:3.2}, as
follows:
\begin{equation}
  \label{eq:4.13}
  2(\Lambda |\alpha)-2(\gamma'|\alpha)
  +2m (k+h\spcheck)=n(\alpha |\alpha)\, .
\end{equation}

We decompose $\alpha \in \fh^{\sigma}(=\fh^{\sigma *})$ with
respect to the orthogonal direct sum decomposition
$\fh^{\sigma}=\CC x+\fhs$:
\begin{equation}
  \label{eq:4.14}
  \alpha =2\alpha (x) x +\alpha^{\natural}\, .
\end{equation}

Next, by Theorem~\ref{th:3.1},  the $W_k (\fg ,\sigma
,\theta/2)$-module $H(M)$ is a Verma module, and its highest weight is
$\lambda =h\delta' +\lambda^{\natural}$, where $h$ is given by
Proposition~\ref{prop:4.1}(a), and (see
Proposition~\ref{prop:4.1}(b)):
\begin{equation}
  \label{eq:4.15}
  \lambda^{\natural}
  =\Lambda^{\natural}-\gamma^{\natural}_{1/2}\, .
\end{equation}

By Proposition~\ref{prop:3.1}, each singular weight
$\hat{\Lambda}-n\hat{\alpha}$ of $M$ satisfying (\ref{eq:4.11})
and such that $\hat{\alpha} \in \hat{\Delta}^{\re}_{++}$, gives
rise to a singular weight of $H(M)$ (which is a Verma module over
$W_k (\fg ,\sigma, \theta /2)$ with highest weight $\lambda$).
This gives rise to a factor of $\det_{\eta}$.  We now
rewrite (\ref{eq:4.13}) in terms of $k$, $h$ and
$\lambda^{\natural}$.

In the case $\alpha (x) =0$, substituting (\ref{eq:4.15}) in
(\ref{eq:4.13}), we obtain (\ref{eq:4.8}).  In the case $\alpha
(x) \neq 0$, we substitute $\Lambda =2\Lambda
(x)x+\lambda^{\natural}+\gamma^{\natural}_{1/2}$ (obtained from
(\ref{eq:4.14}) and (\ref{eq:4.15})) in the formula for $h$ given
by Proposition~\ref{prop:4.1}(a) to obtain:
\begin{eqnarray}
  \label{eq:4.16}
\hspace*{6ex}  h&=&\frac{1}{k+h\spcheck}\left((\Lambda (x) -\gamma'(x)
  -\frac{1}{2} (k+h\spcheck))^2 +\frac{1}{2} |\lambda^{\natural}
  +\gamma^{\natural}_{1/2}-\gamma^{\prime \natural}|^2 \right.\\
\nonumber
  &&\left.-\frac{1}{4} (k+h\spcheck +2\gamma' (x))^2-\frac{1}{2}
  |\gamma^{\prime \natural}|^2\right) +s_{\fg}+s_{\gh}\, .
\end{eqnarray}
Substituting (\ref{eq:4.14}) and (\ref{eq:4.15}) in (\ref{eq:4.13}), we obtain:
\begin{displaymath}
  2\alpha (x) \Lambda (x) =\frac{n}{2}|\alpha |^2
     -(\lambda^{\natural}+\gamma^{\natural}_{1/2} -
        \gamma' |\alpha) -m(k+h\spcheck)\, .
\end{displaymath}
Finally, substituting the obtained expression for $\Lambda (x)$
in (\ref{eq:4.16}) and using Proposition~\ref{prop:4.2}(a), we get (\ref{eq:4.9}) and (\ref{eq:4.10}).

The rest of the proof is the same as in \cite{KW}.

\end{proof}

\begin{remark}
  \label{rem:4.2}

\begin{subremark}
If $\alpha +m\delta \in \hat{\Delta}^{\re}_{++}$ is such that
$(\alpha | \alpha)=0$, the condition (\ref{eq:4.1}) becomes
$(\hat{\Lambda} + \hat{\rho}^{\tw} | \alpha +m\delta)=0$.  Hence in
this case the function $\varphi_{\alpha+m\delta,n}
(k,h,\lambda^{\natural})$
is independent of $n$.  Since $\alpha+m\delta$ is an
odd root, we therefore can simplify the corresponding factor in
the formula for $\det_{\eta}$ (cf.~\cite{KW}, Remark~7.2):
\begin{displaymath}
  \prod_{n \in
    \NN}\varphi_{\alpha+m\delta,n}^{ \tilde{p}(\alpha)^{n+1}
      P_{W,\sigma} (\eta -n\pi(\alpha+m\delta))}
    =\varphi_{\alpha+m\delta ,1}^{ P_{W,\sigma \, ; \,
        \pi(\alpha+m\delta)}    (\eta -\pi(\alpha+m\delta))}\, .
\end{displaymath}
Here $P_{W,\sigma \, ; \, \hat{\alpha}}$ stands for the
partition function of the set $\Delta^+_{W,\sigma} \backslash \{
\hat{\alpha} \}$ (i.e.,~we reduce by $1$ the multiplicity of
$\hat{\alpha}$).
\end{subremark}

\begin{subremark}
If $\alpha + m\delta \in \hat{\Delta}^{\re}_{++}$ is such that $2
(\alpha +m\delta)\in \hat{\Delta}^{\re}_{++}$, and $n \in \NN$, then
condition~(\ref{eq:4.11}) for the pair $\{2 (\alpha +m\delta), n\}$ is
the same as that for the pair $\{\alpha +m\delta, 2n\}$, hence in this
case we have
\begin{displaymath}
  \varphi_{\alpha +m\delta,2n}=\varphi_{2(\alpha
    +m\delta),n}\, ,
\end{displaymath}
and the corresponding factors in $\det_{\eta}$ cancel as in
\cite{KW}, Remark~7.2.
\end{subremark}

\begin{subremark}
In all examples we have:  $\varphi_0 = \varphi_{(-\delta +\theta) /2,0}$,
but we do not know how to prove this in general. We conjecture
that this is always the case, i.e., 
\begin{displaymath}
  \varphi_0 =h- \frac{1}{2(k+h\spcheck)} \left(|
    \lambda^{\natural}+\gamma^{\natural}_{1/2}-\gamma^{\prime
      \natural} |^2
    -\frac{1}{2}(k+\frac{1}{2})^2-|\gamma^{\prime
      \natural}|^2\right)-s_{\fg}-s_{\gh} \, . 
\end{displaymath}

\end{subremark}

%
%

\end{remark}

\section{Examples}
\label{sec:5}

\subsection{Ramond $N=1$ algebra}\qquad
\label{sec:5.1}

Recall that the Neveu-Schwarz vertex algebra is $W_k (spo (2|1),
\, \theta /2)$ \cite{KW}.  It corresponds to the minimal
gradation of $\fg = spo (2|1)$, which looks as follows:
\begin{displaymath}
  \fg = \CC e_{-\theta} \oplus \CC e_{-\theta/2} \oplus
  \CC x \oplus \CC e_{\theta/2} \oplus \CC e_{\theta}\, ,
\end{displaymath}
where $e_{-\theta}=\frac{1}{2}E_{21}$, $e_{-\theta/2}=\frac{1}{2}
(E_{31}-E_{23})$, $x=\frac{1}{2} (E_{11}-E_{22})$,
$e_{\theta/2}=E_{13} +E_{32}$, $e_{\theta} =2E_{12}$, $\fh =\CC
x$, and $\theta \in\fh^*$ is defined by $\theta (x)=1$.  Then
$\Delta_+ =\{ \theta/2 ,\theta \}$.  Choose the invariant bilinear
form $(a|b) =\str ab$.  Then $h\spcheck =3/2$ and $(e_{\theta/2} |
e_{-\theta/2}) = (e_{\theta} |e_{-\theta})=1$, $(x|x)=1/2$.  We
have $x=\theta/2$ under the identification of $\fh$ with $\fh^*$.

We take $f=e_{-\theta}$.  The only non-trivial automorphism
$\sigma$ that fixes $f$ and $x$, also fixes $e=e_{\theta}/2$ and
$\sigma (e_{\pm \theta/2})=-e_{\pm \theta/2}$.  Then we have:
\begin{displaymath}
  \epsilon (\sigma)=1 \, , \, s_{\theta/2} =1/2 \, , \,
    s_{-\theta/2}=1/2\, , \, s_{\theta}=0 \, , \, s_{-\theta}
  =1\, .
\end{displaymath}
Hence we have:
$\gamma'=-\theta/2$, $\gamma_{1/2}=-\theta/8$, $s_{\fg}=-k/4(2k+3)$,$s_{\gh}=
-1/16$.

In this case we have one twisted neutral free fermion $\Phi^{\tw}
(z) =\sum_{n \in \ZZ} \Phi_n$\break $ z^{-n-1/2}$, where $[\Phi_m ,
\Phi_n] =\delta_{m,-n}$ and $\Phi^{\tw}(z)_- =\sum_{n>0} \Phi_n
z^{-n-1/2}$.

The twisted vertex algebra $W_k (\fg ,\sigma ,\theta /2)$ is
strongly generated by the Virasoro field $L^{\tw} (z) =\sum_{n
  \in \ZZ} L^{\tw}_n z^{-n-2}$ and the odd Ramond field
$G^{\tw}(z)=\sum_{n \in \ZZ} G^{\tw}_n z^{-n-3/2}$, so that the
$L_n$ and $G_n$ satisfy the relations of the Ramond $(N=1)$
superalgebra \cite{R} with central charge
\begin{displaymath}
  c(k)=3/2 - 12 \gamma^2 \, , \hbox{ where } \gamma^2 = (k+1)^2/
  (2k+3)\, .
\end{displaymath}
In particular, we have:
\begin{equation}
  \label{eq:5.1}
  [G^{\tw}_0 , G^{\tw}_0] =2L^{\tw}_0 -c(k)/12 \, .
\end{equation}

The free field realization, provided by Theorem~\ref{th:4.1}, of
this algebra is given in terms of a free boson $b(z) =\sum_{n \in
\ZZ} b_n z^{-n-1}$, where $[b_m,b_n] =m\delta_{m,-n}$ and $b(z)_-
=\sum_{n \geq 0} b_n z^{-n-1}$, and the twisted fermion
$\Phi^{\tw} (z)$.  We have:
\begin{eqnarray*}
  L^{\tw} (z) &=& \frac{1}{2}:b(z)^2 :+ \gamma \partial b(z)
     -\frac{1}{2}: \Phi^{\tw}(z)\partial\Phi^{\tw}(z) : -\frac{1}{16}z^{-2}\, , \\
  G^{\tw}(z) &=&  \frac{1}{\sqrt{2}} : \Phi^{\tw} (z) b (z) :
     +\sqrt{2} \gamma \partial \Phi^{\tw} (z) \, .
\end{eqnarray*}

In order to compute the determinant formula for the Ramond
algebra we need the $\sigma$-twisted affinization
$\hat{\fg}^{\tw} =\sum_{m\in \ZZ} \fg_{\bar{0}} t^m + \sum_{m \in
1/2 +\ZZ} \fg_{\bar{1}}t^m + \CC K +\CC D$, where $\fg_{\bar{0}} = \CC
e_{\theta} + \CC x +\CC e_{-\theta}$, $\fg_{\bar{1}}=\CC
e_{\theta/2} + \CC e_{-\theta/2}$.
%
%
%
The set  $\hat{\Delta}^{\re}_{++}$ is a union of two subsets:
\begin{displaymath}
   \{ m\delta +\theta/2 | \, m \in \frac{1}{2}+\ZZ_+ \}
      \hbox{ and }
  \{ m\delta +\theta |\, m \in \ZZ_+ \} \, .
\end{displaymath}
~From (\ref{eq:4.10}) and Remark~\ref{rem:4.2}(b) we obtain that  $\varphi_{m\delta
+\theta/2 ,n} (k,h) =h-h^R_{n,2m+1} (k)$ and $\varphi_{m\delta
+\theta ,n}=h-h^R_{2n,m+1}(k)$, where
\begin{equation}
  \label{eq:5.2}
  h^R_{n,m} (k) = \frac{1}{4(k+\frac{3}{2})}\left(\left(\frac{n}{2} -m
  \left(k+\frac{3}{2}\right)\right)^2 -(k+1)^2\right)+\frac{1}{16}\, .
\end{equation}
It follows from (\ref{eq:5.1}) that the extra factor is equal to
$h-c(k)/24$.

The set of positive even (resp.~odd) roots for $W_k (spo (2|1),\sigma,
\theta /2)$ is $\NN\delta'$ (resp.~$\ZZ_+\delta'$).  Hence $P'_{W,\sigma}(\eta)
=p^R (\eta)$, where $p^R (\eta)$ is defined by the generating
series $\sum_{\eta \in \ZZ_+} p^R (\eta ) q^{\eta}
=\prod^{\infty}_{n=1} \frac{1+q^n}{1-q^n}$.  Hence, by
Theorem~\ref{th:4.2}, we obtain the following determinant formula
for the Ramond algebra, where $\eta \in \NN$ (cf.~\cite{KW}):
\begin{displaymath}
  \det_{\eta} (k,h) = (h-\frac{c(k)}{24})^{p^R(\eta)}
  \prod_{\substack{m,n \in \NN \\ m+n \, \odd}}
  (h-h^R_{n,m} (k))^{2p^R (\eta -\frac{1}{2} mn)}\, .
\end{displaymath}

\subsection{$N=2$ Ramond type sector}\quad
\label{sec:5.2}

Recall that the $N=2$ vertex algebra is $W_k (\sl (2|1), \theta
/2)$.  In this section the Lie superalgebra $\fg = \sl (2|1)$
consists of supertraceless matrices in the superspace
$\CC^{2|1}$, whose even part is $\CC \epsilon_1 +\CC \epsilon_3$
and odd part is $\CC \epsilon_2$, where $C \epsilon_1$, $\epsilon_2$,
$\epsilon_3$  is the standard basis.  We shall work in the
following basis of $\fg$:
\begin{eqnarray*}
&  e_1 = E_{12} \, , \, e_2=E_{23}\, , \, -[e_1,e_2]\, , \,
  f_1 =E_{21} \, , \, f_2=-E_{32}\, , \, [f_1,f_2]\, , \, \\
& h_1 =E_{11}+E_{22}\, , \, h_2=-E_{22}-E_{33}\, .
\end{eqnarray*}
The elements $e_i$, $f_i$, $h_i$ $(i=1,2)$ are the Chevalley
generators of $\fg$ and $\fh =\CC h_1 + \CC h_2$.  The elements
$e_i$, $f_i$ $(i=1,2)$ are all odd elements of $\fg$, both simple
roots $\alpha_i$ $(i=1,2)$, attached to $e_i$, are odd, and
$\Delta_+ = \{ \alpha_1 ,\alpha_2, \theta =\alpha_1 +\alpha_2 \}$.
Since $\fg_{\bar{0}} =\CC [e_1,e_2] +\CC [f_1,f_2] +\fh \simeq
g\ell_2$, there is only one, up to conjugacy, nilpotent element
$f=[f_1,f_2]$, which embeds in the following $\sl_2$-triple:  $\{
e=-\frac{1}{2}[e_1,e_2]\, , \, x=\frac{1}{2}(h_1+h_2)\, , \, f \}$.  The
minimal gradation of $\fg$, defined by $\ad x$, looks as follows:
\begin{displaymath}
  \fg =\CC f \oplus (\CC f_1 + \CC f_2) \oplus \fh
      \oplus (\CC e_1 +\CC e_2) \oplus \CC e, .
\end{displaymath}
The invariant bilinear form on $\fg$ is $(a|b) =\str ab$, and
$h\spcheck =1$.

First consider the Ramond type automorphisms $\sigma_a$ ($-1/2 < a\leq 1/2$),
defined by
$\sigma_a (e_1)=e^{2\pi ia}e_1$, $\sigma_a (f_1)=e^{-2\pi ia}f_1$,
$\sigma_a (e_2)=e^{-2\pi ia}e_2$, $\sigma_a (f_2)=e^{2\pi ia}f_2$.
Then
$\fg (\sigma_a) =\fg$ if $a=1/2$ (resp. $=\fg_{\bar{0}}$ if $a< 1/2$),
and we choose $\fn (\sigma_a)_+ =\CC e_2+\CC f_1 +\CC f \, , \,
\fn (\sigma_a)_- =\CC f_2 +\CC e_1 + \CC e$ (resp. $\CC f$ and $\CC e$) ,
so that in all cases $\epsilon(\sigma_a)=0$, and
\begin{displaymath}
  s_{\alpha_1} =a \, , \, s_{\alpha_2} =-a \, , \,
  s_{\theta}=0 \, , \, s_{-\alpha_1}=1-a \, , \,
  s_{-\alpha_2}=1+a\, , \, s_{-\theta}=1\, .
\end{displaymath}
In this case we have two twisted neutral free fermions
$\Phi^{\tw}_1(z) =\sum_{n \in 1/2+a+\ZZ}$\break $\Phi^1_n z^{-n-1/2}$,
$\Phi^{\tw}_2(z) =\sum_{n \in 1/2-a+\ZZ}\Phi^2_n z^{-n-1/2}$,
where
$[\Phi^i_m\, , \, \Phi^j_n] =$\break $(\delta_{ij}-1)$ $\delta_{m,-n}$,
$\Phi^{\tw}_1 (z)_- =\sum_{n \in 1/2+a+\ZZ_+} \Phi^1_n z^{-n-1/2}$,
$\Phi^{\tw}_2 (z)_- =\sum_{n \in 1/2-a+\ZZ_+}$\break$ \Phi^2_n z^{-n-1/2}$.

The twisted vertex algebra $W_k (\fg ,\sigma_{a}, \theta /2)$ is
strongly generated by the Virasoro field $L^{\tw} (z) =\sum_{n
  \in \ZZ} L^{\tw}_n z^{-n-2}$, the current $J^{\tw}(z)
=\sum_{n \in \ZZ} J^{\tw}_n$ $ z^{-n-1}$ and two odd fields
$G^{\pm,\tw}(z) =\sum_{n \in 1/2 \mp a+\ZZ} G^{\pm ,\tw}_n z^{-n-3/2}$ so
that $L_n$, $J_n$ ($n \in \ZZ$) and $G^{\pm}_n$ $(n \in 1/2 \mp a+\ZZ)$ satisfy the
relations of $N=2$ Ramond type superconformal algebra with central charge
$c(k)=-3(2k+1)$.

The free field realization, provided by Theorem~\ref{th:4.1}, of
this algebra is given in terms of free bosons
$  h_i (z) = \sum_{n \in \ZZ} h^i_n z^{-n-1}$ $(i=1,2)$, where
$[h^i_m,h^j_n] = (k+1) m  (1-\delta_{ij}) \delta_{m,-n}$, and the
twisted neutral free fermions $\Phi^{\tw}_i (z) (i=1,2)$:
\begin{eqnarray*}
  L^{\tw}(z) &=& \frac{1}{k+1}: h_1 (z) h_2 (z) :
  + \frac{1}{2} (:\Phi^{\tw}_1 (z) \partial \Phi^{\tw}_2(z):\\[1ex]
&& +: \Phi^{\tw}_2 (z) \partial \Phi^{\tw}_1 (z) :
   + \partial   (h_1 (z) + h_2 (z) )) +\frac{a^2}{2} z^{-2} \, , \\[1ex]
  J^{\tw}(z) &=& h_1 (z) -h_2 (z) +: \Phi^{\tw}_1 (z)
  \Phi^{\tw}_2 (z): + az^{-1}\, , \\[1ex]
  G^{+,\tw} (z) &=& (-k-1)^{-1/2} (:\Phi^{\tw}_2 (z)h_1(z):+
  (k+1)\partial \Phi^{\tw}_2 (z))\\[1ex]
  G^{-,\tw}(z) &=& (-k-1)^{-1/2} (:\Phi^{\tw}_1 (z) h_2 (z) :+
  (k+1)\partial \Phi^{\tw}_1 (z))\, .
\end{eqnarray*}

The set $\hat{\Delta}_+ =\hat{\Delta}^{\re}_+ \cup
\hat{\Delta}^{\im}_+$ of positive roots of $\hat{\fg}^{\tw}$ is as
follows:  $\hat{\Delta}^{\re}_+ = \{ (m +a) \delta +
\alpha_1 \, , (m-a+1)\delta-\alpha_1\, , (m-a)\delta +\alpha_2 \, , \,
(m+a+1)\delta -\alpha_2 \, , \, m\delta +\theta \, , \,
(m+1) \delta -\theta |\,  m\in \ZZ_+\}$, where all roots have
multiplicity $1$, and $\hat{\Delta}^{\im}_+ =\{ m \delta | \, m
\in \NN \}$, all having multiplicity $2$.  
Next, $\gamma'=-aH$, $\gamma_{1/2}=-aH/2$, where $H=h_1-h_2$,
$s_{\fg}=ka^2/(k+1)$,
$s_{\gh}=-a^2/2$, and the set of roots $\hat{\Delta}^{\re}_{++}$ is as follows:
\begin{displaymath}
  \{ m\delta +\theta | \, m \in \ZZ_+ \} \cup
  \{ m\delta +\alpha_1 | \, m \in a +\ZZ_+ \} \cup
  \{ m\delta +\alpha_2 | \, m \in -a +\ZZ_+ \} \, .
\end{displaymath}
We have:  $\fh_{W,\sigma} =\CC H + \CC L^{\tw}_0$, where $H=h_1-h_2$.  Define
$\alpha \in \fh^*_{W,\sigma}$ by $\alpha (H)=1$, $\alpha
(L^{\tw}_0)=0$.  Then $\Delta^{\re ,+}_{W,\sigma} = \{ m\delta'
-\alpha |\, m \in 1/2 +a+\ZZ_+ \} \cup \{ m\delta' +\alpha | \,
m \in 1/2 -a +\ZZ_+
\}$, all of multiplicity $1$, and $\Delta^{\im ,+}_{W,\sigma}=\{
m\delta' | \, m \in \NN \}$ of multiplicity $2$.  Let $P^a_{N=2}
(\eta)\, , \, \eta \in \fh^*_{W ,\sigma}$, be the corresponding
partition function. Let $h$ and $j$ be the eigenvalues of
$L^{\tw}_0$ and of $J^{\tw}_0$ respectively on the highest weight vector
$v_{\lambda}$.

By Theorem~\ref{th:4.2} and Remark ~\ref{rem:4.2}(a), we
obtain the following formula for $\det_{\eta} (k,h,j)$, conjectured in
\cite{BFK} (cf. \cite{KM}):
\begin{eqnarray*}
  \prod_{m,n \in \NN} \Bigl((k+1) (h-h_{m,n}
  (k,j))\Bigr)^{P^a_{N=2} (\eta-mn\delta')}\\
  \times \prod_{m \in 1/2+a+\ZZ_+} \varphi_{m, -}
  (k,h,j)^{P^a_{N=2; m\delta'-\alpha}(\eta-( m\delta'-\alpha))}\\
\times \prod_{m \in 1/2-a+\ZZ_+} \varphi_{m, +}
  (k,h,j)^{P^a_{N=2;m\delta'+\alpha} (\eta -(m\delta' +\alpha))}\, ,
\end{eqnarray*}
where
\begin{eqnarray*}
h_{m,n} (k,j) &=& \frac{1}{4(k+1)}\Bigl((n-m(k+1) )^2 -
(j-a)^2-(k+1)^2\Bigr) +\frac{a^2}{2}\, ,\\
 \varphi_{m, \pm } (k,h,j) &=& h-(m^2+m)(k+1)\mp (m+\frac{1}{2})(j-a)-
\frac{a^2}{2}\,.
\end{eqnarray*}

\subsection{$N=2$ twisted sector}\quad
\label{sec:5.3}

In this subsection we consider the involution $\sigma =
\sigma_{\tw}$ of $\fg =\sl (2|1)$ defined by:
\begin{displaymath}
  \sigma_{\tw} (e_1)=e_2 \, , \, \sigma_{\tw} (f_1)=f_2\, , \,
  \sigma_{\tw}(h_1)=h_2\, .
\end{displaymath}
Let $e^{(1)}= (e_1+e_2)/\sqrt{2}\, , \, e^{(2)}=(e_1-e_2)/\sqrt{2}
\, , \, f^{(1)} = (f_1 +f_2)/\sqrt{2}\, , \, f^{(2)}=
(f_1-f_2)/\sqrt{2} \, , \, H=h_1-h_2$.  Then
\begin{displaymath}
  \fg (\sigma_{\tw}) =\CC f \oplus \CC f^{(2)} \oplus
  \CC x \oplus \CC e^{(2)} \oplus \CC e \, ,
\end{displaymath}
and the only possible choice for $\fn (\sigma_{\tw})_{\pm}$ is
as follows:
\begin{displaymath}
  \fn (\sigma_{\tw})_+ =\CC f^{(2)} +\CC f \, , \,
  \fn (\sigma_{\tw})_- = \CC e^{(2)} +\CC e\, .
\end{displaymath}
Note that $\fn_{1/2} (\sigma)_+ =\fn_{1/2} (\sigma)'_-=0$ and
$\fg_{1/2}(\sigma)_0 =\CC e^{(2)}$ (see (\ref{eq:2.5})), so that
$\epsilon (\sigma_{\tw})=1$.  Note also that $\fh^{\sigma} =\CC x$, so that
the set $\Delta^{\sigma}$ ($\subset \fh^{\sigma *}$) of non-zero roots of $\fh^{\sigma}$
in $\fg$ is $\Delta^{\sigma} =\{ \pm \theta \, , \, \pm\theta /2 \}$, where
$\theta (x) =1$, the roots $\pm \theta$ (resp.~$\pm \theta /2$)
being of multiplicity~$1$ (resp.~$2$).  Thus the $s_a$ are as
follows:
\begin{displaymath}
  s_H = s_{e^{(2)}} =s_{f^{(2)}} =1/2 \, , \, s_{e^{(1)}} =s_e
     =0 \, , \, s_{f^{(1)}} =s_f =1 \, .
\end{displaymath}
(Note that here $s_a$ depends not only on the root, but also on
the root vector.)

The free field realization of the twisted vertex algebra $W_k
(\fg ,\sigma_{\tw},\theta /2)$, provided by Theorem~\ref{th:4.1},
is given in terms of free neutral fermions $\Phi^{(1)} (z)
=\sum_{n \in 1/2 +\ZZ} \Phi^{(1)}_n z^{-n-1/2}$, $\Phi^{(2)\tw}
(z) =\sum_{n \in \ZZ} \Phi^{(2)}_n z^{-n-1/2}$,  where
$[\Phi^{(i)}_m , \Phi^{(j)}_n] =(-1)^j \delta_{ij}
\delta_{m,-n}$, $\Phi^{(1)}(z)_- =\sum_{n>0} \Phi^{(1)}_n
z^{-n-1/2}$, $\Phi^{(2)\tw}(z)_- =\sum_{n>0}\Phi^{(2)}_n$\break
$z^{-n-1/2}$, and free commuting bosons $x(z) =\sum_{n \in \ZZ}
x_n z^{-n-1}$, $H^{\tw} (z) =\sum_{n \in 1/2 +\ZZ} H_n$ $ z^{-n-1}$,
where $[x_m,x_n] =\frac{1}{2}(k+1) m\delta_{m,-n}$, $[H_m,H_n]
=$\break $-2(k+1)m\delta_{m,-n}$, $x(z)_- =\sum_{n \geq 0} x_n z^{-n-1}$
and $H^{\tw} (z)_- =\sum_{n>0}H_n z^{-n-1}$:
\begin{eqnarray*}
  L^{\tw} (z) &=& \frac{1}{k+1}
     \left( :x(z)^2 :- \frac{1}{4}:H^{\tw}(z)^2:\right)
     +\frac{1}{2} \left( :\Phi^{(1)} (z) \partial
       \Phi^{(1)}(z):\right. \\
     && \left. -: \Phi^{(2),\tw} (z) \partial \Phi^{(2),\tw}(z):\right)
    +\partial x(z) \, , \\[1ex]
  J^{\tw}(z) &=& H^{\tw}(z)-:\Phi^{(1)}(z) \Phi^{(2),\tw}(z):\, ,
  \\[1ex]
  G^{(1),\tw} (z) &=& (-k-1)^{-1/2} \left( :\Phi^{(1)} (z) x(z) :
       -\frac{1}{2}: \Phi^{(2),\tw} (z) H^{\tw}(z):\right. \\
       && \left.+(k+1) \partial \Phi^{(1)}(z) \right) \, , \\[1ex]
  G^{(2),\tw} (z) &=& (-k-1)^{-1/2}\left( :\Phi^{(2),\tw}(z)
    x(z) :+\frac{1}{2}:\Phi^{(1)} (z) H^{\tw} (z) :\right. \\
  && \left.  -(k+1)    \partial \Phi^{(2),\tw} (z) \right) \, ,
\end{eqnarray*}
where $G^{(1),\tw}=\frac{1}{\sqrt{2}} (G^{+,\tw}+G^{-,\tw})$,
$G^{(2),\tw} =\frac{1}{\sqrt{2}} (G^{+,\tw}-G^{-,\tw})$.

Furthermore, in this case the set $\hat{\Delta}_+ =
\hat{\Delta}^{\re}_+ \cup \hat{\Delta}^{\im}_+$ of positive roots
of $\hat{\fg}^{\tw}$ is as follows:  $\hat{\Delta}^{\re}_+ =\{ m
\delta +\theta /2$, $(m+1)\delta -\theta /2$, $m\delta +\theta$,
$(m+1)\delta -\theta |\, m \in \ZZ_+ \} \cup \{ m\delta \pm
\theta /2|\, m \in \frac{1}{2}+\ZZ_+ \}$, $\hat{\Delta}^{\im}_+ =
\{ m\delta |\, m \in \frac{1}{2} \NN \}$, all having
multiplicity~$1$. Note also that the roots $m\delta \pm \theta/2$
are odd and all the other roots are even.  
We have: $\fh^{\natural}=0$,
$s_{\fg}=-k/16(k+1)$, $s_{\gh}=-1/16$, and
\begin{displaymath}
  \hat{\Delta}^{\re}_{++} =\{ m \delta +\theta |\, m \in \ZZ_+\}
  \cup \{ m\delta +\theta /2 |\, m \in \frac{1}{2}\ZZ_+ \}\, ,
\end{displaymath}
all of multiplicity~$1$.  From (\ref{eq:4.10}) and
Remark~\ref{rem:4.2}(b) we obtain that\break $\varphi_{m\delta +\theta
  /2 ,n} (k,h)=h-h^{\tw}_{n,2m+1}(k)$ and $\varphi_{m\delta
  +\theta ,n} (k,h) = h-h^{\tw}_{2n,m+1} (k)$, where
\begin{equation}
  \label{eq:5.3}
  h^{\tw}_{n,m}(k) = \frac{1}{4(k+1)} \left(\left( \frac{n}{2}-m
    (k+1)\right)^2-(k+1)^2\right) +\frac{1}{8}\, .
\end{equation}
It is easy to compute that $[G^{(2)}_0,G^{(2)}_0]=-(L_0-c(k)/24)$,  hence
the extra factor equals
$\varphi_{(\theta -\delta)/2,0} (h,k)=h-c(k)/24 =h+(2k+1)/8$.

The set of positive even (resp. odd) roots for $W_k (\sl (2|1),
\sigma^{\tw}, \theta /2)$ is $\frac{1}{2}\NN \delta'$
(resp.~$\frac{1}{2} \ZZ_+ \delta'$), all of multiplicity~1.
Hence $P'_{W,\sigma^{\tw}} (\eta) =p^{\tw}(\eta )$, where
$p^{\tw} (\eta )$ is defined by the generating series $\displaystyle{\sum_{\eta
  \in \frac{1}{2}\ZZ_+} p^{\tw} (\eta) q^{\eta}
=\prod^{\infty}_{n=1} \frac{1+q^{n/2}}{1-q^{n/2}}}$.  Hence by
Theorem~\ref{eq:4.2}, we obtain the following determinant formula
for the $N=2$ twisted superconformal algebra, conjectured in \cite{BFK}:
\begin{displaymath}
\det_{\eta} (k,h)=(h+\frac{2k+1}{8})^{p^{\tw}(\eta)}
\prod_{\substack{m,n \in \NN\\ n\, \odd}}
(h-h^{\tw}_{n,m} (k))^{2p^{\tw}(\eta -\frac{1}{2}mn)}\, .
\end{displaymath}

\subsection{$N=4$ Ramond type sector}\quad
\label{sec:5.4}

Recall that the $N=4$ vertex algebra is $W_k (\fg ,\theta/2)$,
where $\fg =\sl (2|2)/\CC I$.  We shall use the same basis of
$\fg$ and keep the same notation as in \cite{KW}, Section~8.4.
In particular, the simple roots are $\alpha_1$, $\alpha_2$,
$\alpha_3$, where $\alpha_1$ and $\alpha_3$ are odd and
$\alpha_2$ is even, all the non-zero scalar products between them
being $(\alpha_1|\alpha_2) = (\alpha_2|\alpha_3)=1$,
$(\alpha_2|\alpha_2)=-2$.  The dual Coxeter number $h\spcheck =0$.

Consider the Ramond type automorphisms $\sigma =\sigma_{a,b}$ of
$\fg$, where $-1/2 <a$, $b \leq 1/2$, defined by $\sigma
(e_1)=e^{2\pi ia} e_1$, $\sigma (e_2)=e^{-2\pi i (a+b)} e_2$,
$\sigma (e_3)=e^{2\pi ib}e_3$, $\sigma (h_i)=h_i$.   Note that
$\epsilon (\sigma_{a,b})=0$.  We consider first the case when
$a+b>0$.  Then we have the following possibilities for $\fn
(\sigma)_{\pm}$:

\romanparenlist
\begin{enumerate}
\item 
$a,b \neq 1/2$: $\fn (\sigma)_-=\CC e$, where $e=e_{12 3}$,
$\fn(\sigma)_+=\CC f$, where $f=f_{123}$;

\item 
$a=1/2$, $b \neq 1/2$: $\fn(\sigma)_- =\CC e +\CC e_1 + \CC
f_{23}$, $\fn (\sigma)_+ =\CC f +\CC e_{23}$; $+\CC f_1$;

\item 
$a \neq 1/2$, $b=1/2$: $\fn (\sigma)_- = \CC e+\CC e_3 +\CC
f_{12}$, $\fn (\sigma)_+ = \CC f + \CC e_{12}+\CC f_3$;

\item 
$a=b=1/2$: $\fn (\sigma)_- =$ span $\{ e,e_1, e_3, f_{12}, f_{23},
f_2 \}$, $\fn (\sigma)_+ =$\break span $\{ f,e_{12}, e_{23}, e_2, f_1,
f_3 \}$.
\end{enumerate}

In these four cases the $s_{\alpha}$ are  as follows:
\begin{displaymath}
      s_{\alpha_1} =a,s_{\alpha_2}=1-a-b, s_{\alpha_3}=b,
        s_{\alpha_1 + \alpha_2}=-b, s_{\alpha_2 +\alpha_3}=-a ,
        s_{\theta}=0 \, ,
\end{displaymath}
where $\theta =\alpha_1 +\alpha_2+\alpha_3$, and, as usual,
$s_{-\alpha}=1-s_{\alpha}$.  Consequently, we have:
\begin{eqnarray*}
\hat{\Delta}^{\re}_{++}& =&\{ (m+a)\delta +\alpha_1 \, , \,  (m+b) \delta
+\alpha_3 \, ,\,  (m-b)\delta +\alpha_1+\alpha_2 \, , \\
&& (m-a)\delta +\alpha_2 +\alpha_3 \,, \,
 (m+1-a-b)\delta +\alpha_2\, , \, (m+a+b)\delta
-\alpha_2 \,,\\
 && m\delta +\theta |\, m \in \ZZ_+ \}\, .
\end{eqnarray*}
Next $\gamma^{\prime\natural}=\frac{1}{2}\alpha_2$, 
$\gamma^{\natural}_{1/2}=\frac{a+b}{2}\alpha_2$, $s_{\fg}+s_{\gh}=-ab +(a+b)/2$.

We have:  $\fh^{\natural} =\CC h_2$, hence $\fh_{W,\sigma} = \CC h_2+\CC
L^{\tw}_0$.  Define $\alpha \in \fh^*_{W,\sigma}$ by $\alpha
(h_2)=2$, $\alpha (L^{\tw}_0)=0$.  Then
\begin{eqnarray*}
  \Delta^+_{W,\sigma} = \{
  (m+\frac{1}{2}+a)\delta'-\frac{\alpha}{2} , 
  (m+\frac{1}{2}+b)\delta'-\frac{\alpha}{2} , 
 (m+\frac{1}{2}-b)\delta'+\frac{\alpha}{2} ,\\
  (m\!+\!\frac{1}{2}\!-\!a)\delta'\!+\!\frac{\alpha}{2} , 
  (m\!+\!1-a-b)\delta'\!+\!\alpha , 
 (m\!+\!a\!+\!b) \delta' \!-\!\alpha'  ,  (m\!+\!1)\delta' |\, m \in \ZZ_+ \}
\end{eqnarray*}
is the set of positive roots of $W(\fg ,\sigma_{a,b},\theta /2)$,
all having multiplicity~$1$, except for $m\delta'$ which have
multiplicity~$2$.  We have:  $\alpha^{\natural}_1
=\alpha^{\natural}_3=-\alpha /2$, $\alpha^{\natural}_2=\alpha$.

Let $P^{a,b}_{N=4} (\eta )$ be the corresponding partition
function.  Let $h$ and $j$ be the eigenvalues of $L^{\tw}_0$ and
$J^{\{ h_2 \},\tw}$ on $v_{\lambda}$, so that $\lambda^{\natural}
=\frac{j}{2}\alpha $.  Formulas (\ref{eq:4.8}) ---
(\ref{eq:4.10}) give the following factors of the determinant (we
introduce a simplifying notation consistent with the map $\pi$):
$  \varphi_{m,n} := k\varphi_{(m-1)\delta +\theta ,n}$,
  $\varphi_{m,-\alpha /2}:= \varphi_{m\delta +\alpha_1,1}
    = \varphi_{m\delta +\alpha_3,1}$,
  $\varphi_{m,\alpha/2} := \varphi_{m\delta +\alpha_1+\alpha_2,1}
    =\varphi_{m\delta +\alpha_2 +\alpha_3 ,1}$,
  $\varphi_{m,n,\pm\alpha} =\varphi_{m\delta \pm\alpha_2 ,n}$,
  where:
  \begin{eqnarray*}
    \varphi_{m,n} (k,h,j)\!\!\!\!\! &=&\!\!\!\!\! 4kh\!-\!(n\!-\!mk)^2\!\! +\! (a\!+\!b\!+\!j\!-\!1)^2\!
       \!+\!k(k\!+\!1)\!+\!k( 2a\!-\!1)(2b\!-\!1)  ,\\
    \varphi_{m,\pm \alpha/2} (k,h,j)\!\!\!\!\!
      &=&\!\!\!\!\! h-\Big(m+\frac{1}{2}\Big)^2 k\pm
      \Big( m+\frac{1}{2}\Big) (a+b+j-1)+\frac{k+1}{4}\\
     &&\!\!\!\! +\Big( a-\frac{1}{2}\Big)\Big( b-\frac{1}{2}\Big)\, ,\\
    \varphi_{m,n,\pm \alpha} (k,j )\!\!\!\! &=&\!\!\!\! mk \mp (a+b+j-1)+n\, .
  \end{eqnarray*}
By Theorem~\ref{th:4.2} and Remark~\ref{rem:4.2}(a), we obtain
the following formula for $\det_{\eta} (k,h,j)$ in the case 
$a+b>0$:
\begin{align*}
 \lefteqn{ \hspace{-5ex} \prod_{m,n \in \NN} \varphi_{m,n}
          (k,h,j)^{P^{a,b}_{N=4} (\eta -mn \delta')}}\\
 & \times \prod_{m\in \frac{1}{2}+\{ a,b \} +\ZZ_+}
          \varphi_{m,-\alpha/2}
          (k,h,j)^{P^{a,b}_{N=4; m\delta'-\alpha/2}
          (\eta -(m\delta' -\alpha/2))}\\
  & \times \prod_{m\in \frac{1}{2}-\{a,b\}+\ZZ_+}
         \varphi_{m,\alpha/2}
         (k,h,j)^{P^{a,b}_{N=4 ; m\delta' +\alpha/2}
            (\eta -(m\delta' +\alpha /2))}\\
   &\times \prod_{\substack{m\in a +b+\ZZ_+\\ n\in \NN}}
         \varphi_{m,n,-\alpha}
           (k,j)^{P^{a,b}_{N=4}(\eta -n(m\delta'-\alpha))}\\
  & \times \prod_{\substack{m \in -a-b+\NN \\ n\in \NN}}
        \varphi_{m,n,\alpha}
          (k,j)^{P^{a,b}_{N=4}(\eta -n(m\delta'+\alpha))}\, .
\end{align*}
The case $a+b \leq 0$ is treated in the same fashion.   The
$s_{\alpha}$'s in this case are the same as in the case $a+b>0$,
except for $s_{\alpha_2}=-a-b$.  After the
calculation, it turns out that the determinant formula in this
case can be obtained from the above determinant formula by
replacing $a$ by $a+1$ and $b$ by $b+1$ in all factors and by changing
the range of $m$ in the last two factors by exchanging $\ZZ_+$
and $\NN$.  Some cases of this determinant formula were
conjectured in \cite{KR}.

We shall omit the free field realization of the Ramond type
sector of $N=4$ and other remaining superconformal algebras as
being quite long.  On the other hand, as in the simplest cases of
$N=1$ and $2$, they are straightforward applications of Theorem~\ref{th:4.1}.

\subsection{$N=3$ Ramond type sector}\quad
\label{sec:5.5}

Recall that the $N=3$ vertex algebra is $W_k (\fg , \theta/2)$,
where $\fg =spo (2|3)$.  (To get the ``linear'' $N=3$
superconformal algebra one needs to tensor the above vertex
algebra with one free fermion, and the results of this section
can easily be extended to the latter case as in \cite{KW}.)  We
shall keep the notation of \cite{KW}, Section~8.5.  In
particular, the simple roots are $\alpha_1$ and $\alpha_2$, where
$\alpha_1$ is odd and $\alpha_2$ is even,  the  scalar
products between them being  $(\alpha_1 |\alpha_1)=0$, $(\alpha_1 |\alpha_2)=1/2$,
$(\alpha_2|\alpha_2)=-1/2$. Since $\theta=2\alpha_1+2\alpha_2$, we have:
$\alpha^{\natural}_2=-\alpha^{\natural}_1=\alpha_2$,
Recall also that $S_0 =\{ \pm \alpha_2 \}$, $S_{1/2}=\{ \alpha_1, \theta/2 ,
\alpha_1+2\alpha_2 \}$,
$S_1=\{ \theta \}$. The dual Coxeter number $h\spcheck
=1/2$.

Consider the Ramond type automorphisms $\sigma =\sigma_{a,b}$ of
$\fg$ defined by $\sigma (e_{10})=e^{2\pi ia}e_{10}$, $\sigma
(e_{01})=e^{2\pi ib}e_{01}$, $\sigma |_{\fh}=1$, where $a,b \in
\RR$ are such that $a+b \in \frac{1}{2}\ZZ$.  We consider the
following three cases:
\begin{eqnarray*}
&&\hbox{I (resp.~II):  } a=-b, -1/2 <a\leq 0 \,\,
\hbox{  (resp. } a=-b\, ,0<a\leq 1/2)\, ,\\
&&\hbox{III:  }  a+b=1/2, -1/2 <a \leq 1/2\, .
\end{eqnarray*}
Note that $\epsilon (\sigma)=0$ in cases~I and~II, and $\epsilon
(\sigma)=1$ in case~III, when $\fg_{1/2}(\sigma)_0 =\CC e_{11}$
in (\ref{eq:2.5}).  We have the following possibilities for $\fn
(\sigma)_{\pm}$:

\begin{list}{}{}
\item I, II , $a \neq 0,1/2 : \fn (\sigma )_- =\CC e_{22}$, $\fn
  (\sigma)_+=\CC f_{22}$,

\item I, II, $a=0 : \fn (\sigma)_- =\CC e_{22}+\CC f_{01}$,
  $\fn(\sigma)_+ =\CC f_{22}+\CC e_{01}$,

\item I, II, $a=1/2 : \fn (\sigma)_- =$ span $\{ e_{22}, e_{10},
  f_{12}\}$, $\fn (\sigma)_+ =$ span $ \{ f_{22} ,e_{12}, f_{10}\}$,

\item III, $a \neq 1/2 : \fn (\sigma)_- =\CC e_{22}+\CC e_{11}$,
  $\fn (\sigma)_+ =\CC f_{22}+\CC f_{11}$,

\item III, $a=1/2 : \fn (\sigma)_- =$ span $\{ e_{22}, e_{11},
  f_{12}, f_{01}, e_{10}\}$,\\
      \hspace*{7.5ex} $\fn (\sigma)_+=
      $ span $\{ f_{22},  f_{11}, e_{12}, e_{01}, f_{10} \}$.

\end{list}

In these cases the $s_{\alpha}$ are as follows  (up to the relation~(\ref{eq:3.3})):

\Romanlist
\begin{enumerate}{}{}
 \item 
   : $s_{\alpha_1} =a$, $s_{\alpha_2}=-a$, $s_{\theta /2} =0$,
  $s_{\alpha_1 + 2\alpha_2}=-a$, $ s_{\theta}=0$;

\item 
  : the same as in I, except for $s_{\alpha_2}=1-a$;

\item 
  : $s_{\alpha_1}=a$, $s_{\alpha_2}=1/2-a$, $s_{\theta
    /2}=1/2$, $s_{\alpha_1+2\alpha_2}=-a$, $s_{\theta}=0$.

\end{enumerate}

One finds that in these three cases:

\Romanlist
\begin{enumerate}{}{}
 \item 
   : $\gamma^{\prime\natural}=(a-\frac{1}{2}) \alpha_2$,
    $\gamma_{1/2}^{\natural} =a\alpha_2$, $s_{\fg}+ s_{\gh}
     =\frac{a(1-a)}{4k+2}+\frac{a}{2}$;
\item 
 : $\gamma^{\prime\natural} = (a+\frac{1}{2})\alpha_2$,
  $\gamma_{1/2}^{\natural}=a\alpha_2$, $s_{\fg} +s_{\gh}
  =-\frac{a(a+1)}{4k+2}+\frac{a}{2}$;

\item 
  : $\gamma^{\prime\natural} =a\alpha_2$,
  $\gamma_{1/2}^{\natural}=a\alpha_2$,
  $s_{\fg}+s_{\gh}=-\frac{a^2}{4k+2}-\frac{1}{16}$.

\end{enumerate}

Consequently we have in these cases $(m \in \ZZ_+)$:
\Romanlist
\begin{enumerate}{}{}
 \item 
   : $\hat{\Delta}^{\re}_{++} =\{ (m-a)\delta +\alpha_2$,
   $(m+1+a) \delta -\alpha_2$, $(m+a)\delta +\alpha_1$,
   $(m-a)\delta +\alpha_1+2\alpha_2$, $m\delta+\theta /2$,
   $m\delta +\theta \}$,

\item 
  :  $\hat{\Delta}^{\re}_{++} =\{ (m+1-a)\delta +\alpha_2$,
  $(m+a)\delta -\alpha_2$, $(m+a)\delta +\alpha_1$,  $(m-a)\delta
  +\alpha_1+2\alpha_2$,  $m\delta+\theta /2$, $m\delta +\theta \}$,

\item 
  :  $\hat{\Delta}^{\re}_{++} =\{ (m+1/2-a)\delta +\alpha_2$,
  $(m+1/2+a)\delta -\alpha_2$, $(m+a)\delta +\alpha_1$,
  $(m-a)\delta +\alpha_1+2\alpha_2$, $(m+1/2)\delta +\theta /2$,
  $m\delta +\theta \}$.

\end{enumerate}

We have:  $\fh^{\natural}=\CC \alpha_2$, hence $\fh_{W,\sigma}=\CC \alpha_2 +\CC
L^{\tw}_0$.  Define $\alpha \in \fh^*_{W,\sigma}$ by $\alpha
=\alpha_2|_{\fh^{\natural}}$, $\alpha (L^{\tw}_0)=0$.  Then we have in the
three cases ($m\in \ZZ_+$):

\Romanlist
\begin{enumerate}{}{}
 \item 
   : $\Delta^+_{W,\sigma} =\{ (m-a)\delta'+\alpha$
, $(m+1 +a) \delta' -\alpha$, $(m+1/2 +a)\delta' -\alpha$,
$(m+\frac{1}{2}-a)\delta' +\alpha$, $(m+1/2)\delta'$,
$(m+1)\delta' \}$;

\item 
  :  $\Delta^+_{W,\sigma} =\{ (m+1-a)\delta' +\alpha$,
  $(m+a)\delta' -\alpha$, $(m+1/2 +a)\delta'-\alpha$,
  $(m+1/2-a)\delta'+\alpha$, $(m+1/2)\delta'$, $(m+1)\delta' \}$;

\item 
  :  $\Delta^+_{W,\sigma} =\{ (m+1/2-a)\delta' +\alpha$,
  $(m+1/2+a)\delta'-\alpha$, $(m+1)\delta'\}$.

\end{enumerate}
The multiplicities of these positive roots of $W(\fg ,\sigma
,\theta /2)$ are $1$, except for the following cases:  $\mult
(m+1)\delta'=2$ in cases~I and II,
$\mult (m+1/2 \mp a)\delta'\pm\alpha =2$
and $\mult (m+1)\delta'=3$ in case~III
$(m\in \ZZ_+)$. Note, however, that in case III we have, in fact,
one even root and one odd root equal
$(m+1/2 \mp a)\delta'\pm\alpha $, each having multiplicity $1$,
and an even (resp. odd) root $(m+1)\delta'$ of multiplicity $2$ (resp. $1$).

We have: $\alpha_2^{\natural}=-\alpha_1^{\natural}=\alpha$.
Note that $0$ is a (odd) root of $\Delta^+_{W,\sigma}$ only in case~III.

Let $P^{a,b}_{N=3} (\eta)$ be the corresponding partition
function.  Let $h$ and $j$ be the respective eigenvalues of $L^{\tw}_0$ and
$J^{\{-4\alpha_2\} ,\tw}_0$ on $v_{\lambda}$, so that $\lambda^{\natural}=\frac{j}{2}\alpha$.

Introduce the following notations for the factors of the
determinant:  $\varphi_{m,n}=\varphi_{(m-1)\delta +\theta}$,
$\varphi_{m,\alpha}=\varphi_{m\delta +\alpha_1 +2\alpha_2,1}$,
$\varphi_{m,-\alpha}=\varphi_{m \delta +\alpha_1,1}$,
$\varphi_{m,n,\pm \alpha}=\varphi_{m\delta \pm \alpha_2,n}$.
Formulas (\ref{eq:4.8}) --- (\ref{eq:4.10}) give the following
expressions in case~I:
\begin{gather*}
  \varphi_{m,n}
  (k,h,j)=h-\frac{1}{4k+2}\Big(\big(m(k+\frac{1}{2})-\frac{n}{2}\big)^2-
  \frac{(j+1)^2}{4}\Big) +\frac{1}{4}(
  k+\frac{3}{2})+\frac{a}{2}\, ,\\ 
   \varphi_{m,\pm \alpha} (k,h,j)=h-( m+\frac{1}{2})^2
       (k+\frac{1}{2}) \pm \frac{1}{2}(m+\frac{1}{2}) (j+1)
       +\frac{1}{4} ( k+\frac{3}{2}) +\frac{a}{2}\, ,\\
  \lefteqn{\hspace{-2.35in}  \varphi_{m,n,\pm \alpha} (k,j)=m ( k+\frac{1}{2})
        +\frac{n}{4} \mp \frac{j+1}{4}\, .}\hfill
\end{gather*}
By Theorem~\ref{th:4.2} and Remarks~4.2(a) and (b) we obtain the
following formula for $\det_{\eta} (k,h,j)$ in case~I (a special case of this formula was 
conjectured in \cite{KMR} and partially proved in \cite{M}):
\begin{gather*}
  \prod_{m,n \in \NN} ( k+\frac{1}{2})^{P^a_{N =3} (\eta -mn\delta')} 
  \prod_{\substack{m,n \in \NN \\ m+n \even}}
        \varphi_{m,n} (k,h,j)^{P^a_{N=3} (\eta - \frac{1}{2}mn\delta')}\\
   \times \prod_{m\in \mp a +\frac{1}{2}+\ZZ_+} \varphi_{m,\pm \alpha}
     (k,h,j)^{P^a_{N=3; m\delta' \pm \alpha} (\eta -(m\delta' \pm \alpha))}\\
  \times  \prod_{\substack{m\in -a+\ZZ_+\\n \in \NN}} \!\!
      \varphi_{m,n,\alpha}(k,j)^{P^a_{N=3} (\eta -n
        (m\delta'+\alpha))}   \!\!\!
  \prod_{\substack{m\in a+\NN\\ n\in \NN}}  \!
      \varphi_{m,n,-\alpha}(k,j)^{P^a_{N=3}
      (\eta -n (m\delta'- \alpha))}\, .
\end{gather*}

In case~II the determinant formula is similar.  It can be
obtained from the above formula by replacing $j+1$ by $j-1$ and
$a$ by $-a$ in all factors and by changing the range of $m$ in
the last two factors by exchanging $\ZZ_+$ and $\NN$.

In case~III we have:
\begin{gather*}
  \varphi_{m,n} (k,h,j)=h-\frac{1}{4k+2}\Big(
    \big(m\big(k+\frac{1}{2}\big)-\frac{n}{2}\big)^2
    -\frac{j^2}{4}\Big) +\frac{k}{4}+\frac{3}{16}\, ,\\
  \varphi_{m,\pm \alpha} (k,h,j) =h-\big( m+\frac{1}{2}\big)^2
      \big( k+\frac{1}{2}\big) \pm \frac{j}{2}\big(m+\frac{1}{2}
        \big) +\frac{k}{4}+\frac{3}{16}\, ,\\
    \lefteqn{\hspace{-2.2in}    \varphi_{m,n,\pm \alpha}(k,j)= m\big(
          k+\frac{1}{2}\big) +\frac{n}{4}\mp \frac{j}{4}\, .}\hfill
\end{gather*}
The extra factor is $\varphi_0=h+\frac{1}{16} ( 4k+3
  +\frac{j^2}{k+1/2})$, which is computed, using formula (\ref{eq:4.7})
(in this case $h\spcheck_0=-1/2$).

By Theorem~\ref{th:4.2} and Remarks~4.1 and
\ref{rem:4.2}(a) and (b) we obtain the following formula for
$\det_{\eta}(k,h,j)$ in case~III:
\begin{gather*}
   \Big( k+\frac{1}{2}\Big)^{\sum_{m,n \in \NN} P^a_{N=3}
     (\eta-mn\delta')+\sum_{m \in \NN} P^a_{N=3;m\delta'}
     (\eta -m\delta')}\\
\lefteqn{\hspace{-2.35in}\times \Big(h+\frac{1}{16} \big(  
    4k+3+\frac{j^2}{k+1/2}
    \big)\Big)^{P^{\prime a}_{N=3}(\eta)}
    \prod_{\substack{m,n \in \NN \\ m+n \,\odd}}
      \varphi_{m,n}(k,h,j)^{P^a_{N=3}(\eta -\frac{1}{2}mn\delta')}}\\
\lefteqn{\hspace{-2in}  \times \prod_{m \in \frac{1}{2}\mp a+\ZZ_+}
      \varphi_{m, \pm \alpha}(k,h,j)^{P^a_{N=3 ; m\delta' \pm
          \alpha} (\eta -(m\delta'\pm \alpha))}}\\
\lefteqn{\hspace{-2in}   \times \prod_{\substack{m\in \frac{1}{2} \mp a+\ZZ_+\\ n\in
       \NN}}  \varphi_{m,n,\pm \alpha}(k,j)^{P^a_{N=3}(\eta
     -n(m\delta'\pm\alpha))}
\, .}
\end{gather*}

\subsection{Big $N=4$ Ramond type sector}\quad
\label{sec:5.6}

Recall that the big $N=4$ vertex algebra is $W_k (\fg ,\theta
/2)$, where $\fg =D(2,\!1;a)$.  (To get the ``linear'' $N\!\!=\!4$
superconformal algebra (\cite{KL},\cite{S},\cite{STP}) one needs to tensor the
above vertex algebra with four free fermions and one free boson
\cite{GS}, and the results of this and the next section can
easily be extended to the latter case as in \cite{KW}.)  We shall
keep the notation of \cite{KW}, Section~8.6.  In particular, the
simple roots are $\alpha_1$, $\alpha_2$, $\alpha_3$, where
$\alpha_1$ and $\alpha_3$ are even, and $\alpha_2$ is odd, the
non-zero scalar products between them being $(a\neq 0 ,-1)$:
\begin{displaymath}
  (\alpha_1 |\alpha_2)=\frac{1}{a+1}\, , \, (\alpha_2|\alpha_3)=
  \frac{a}{a+1}\, , \, (\alpha_1|\alpha_1)=-\frac{2}{a+1}\, , \, 
  (\alpha_3 |\alpha_3)=-\frac{2a}{a+1}\, .
\end{displaymath}
We shall slightly simplify notation of \cite{KW} by letting $e_1
=e_{100}$, $e_2=e_{010}$, $e_3=e_{001}$, $f_1=f_{100}$,
$f_2=f_{010}$, $f_3=f_{001}$, $e=e_{121}$, $f=f_{121}$.

In this subsection we consider the Ramond type automorphisms $\sigma =\sigma_{\mu,\nu}$
of $\fg$ defined by $\sigma (e_1)=e^{2\pi i\mu} e_1$, $\sigma
(e_2)=e^{-\pi i (\mu +\nu)} e_2$, $\sigma (e_3)=e^{2\pi i \nu}
e_3$, $\sigma |_{\fh}=1$, where $\mu ,\nu \in \RR$ are such that
$-1\leq \mu \pm \nu <1$.  We consider separately the following
four cases:  $(++): \mu ,\nu \geq 0$; $(-+): \mu <0  ,\nu \geq 0
$; $(+-): \mu \geq 0, \nu <0$; $(--): \mu ,\nu <0$.  In all
cases, $\epsilon (\sigma)=0$  and $\fh^{\sigma}=\fh$.  
Since $\theta =\alpha_1+2\alpha_2+\alpha_3$, we have:  $\fhs =\CC
\alpha +\CC \alpha'$, where $\alpha :=\alpha_1 |_{\fhs}
=\alpha^{\natural}_1$, $\alpha' :=\alpha_3
|_{\fhs}=\alpha^{\natural}_3$, and $\alpha^{\natural}_2=-(\alpha
+\alpha')/2$.  Recall also that $S_0=\{ \pm \alpha_1, \pm
\alpha_3 \}$, $S_{1/2}=\{ \alpha_2,\alpha_1+\alpha_2 \, , \,
\alpha_2+\alpha_3 \, , \, \alpha_1+\alpha_2+\alpha_3 \}$, $S_1
=\{ \theta \}$.  The dual Coxeter number $h\spcheck =0$.

We have the following
possibilities for $\fn (\sigma)_{\pm}$:

\romanparenlist
\begin{enumerate}
\item 
$\mu =-1$, $\nu=0$: $\fn (\sigma)_- = $ span $ \{ e
,e_2,e_{011},f_1,f_3,f_{111},f_{110}\} $,\hfill\break
$\fn (\sigma)_+ =$ span $\{ f,f_2,f_{011}, e_1,e_3,e_{111},e_{110}
\}$;

\item 
$\mu +\nu\!=\!-1$, $\nu \neq 0$:~$\fn (\sigma)_- =$ span $\{
e,e_2,f_{111}\}$, $\fn (\sigma)_+ \!=$\! span $\{ f, e_{111},f_2 \}$;

\item 
$\mu -\nu\!=\!-1$, $\nu \neq 0$:~$\fn (\sigma)_- \!=$\! span$\{
e,e_{011},\!f_{110}\}$, $\fn (\sigma)_+ \!= \!$ span$\{
f,e_{110},\!f_{011}\!\}$;

\item 
$\mu =\nu=0$: $\fn (\sigma)_- = $ span $\{ e,f_1,f_3 \}$, $\fn
(\sigma)_+ =$ span $\{ f,e_1,e_3 \}$;

\item 
$\mu =0$, $\nu \neq 0$: $\fn (\sigma)_- =\CC e+\CC f_1$, $\fn
(\sigma)_+ =\CC f+\CC e_1 $;

\item 
$\mu \neq 0$, $-1$, $\nu =0$: $\fn (\sigma)_- =\CC e+\CC f_3$,
$\fn (\sigma)_+ = \CC f+\CC e_3$;

\item 
in all other cases:  $\fn (\sigma)_- =\CC e$, $\fn (\sigma)_+=\CC f$.
\end{enumerate}

The $s_{\alpha}$ are as follows:
$s_{\theta}=0$,$s_{\alpha_2}=-\frac{\mu +\nu}{2}$, $s_{\alpha_1 +\alpha_2+\alpha_3} =
\frac{\mu +\nu}{2}$, $s_{\alpha_1+\alpha_2} =\frac{\mu -\nu}{2}$,  
$s_{\alpha_2+\alpha_3} =-\frac{\mu -\nu}{2}$
in all cases; the remaining $s_{\alpha}$ (up to the relation~(\ref{eq:3.3})) are:
$s_{\alpha_1}= \mu$ in cases $(++)$ and $(+-)$, $s_{\alpha_1}=1+\mu$  in cases $(-+)$ and  $(--)$;
$s_{\alpha_3}= \nu$ in cases $(++)$ and $(-+)$, $s_{\alpha_3} =1+\nu$ in cases $(+-)$ and $(--)$.

Using these data one finds that
$\gamma^{\natural}_{1/2}=-\frac{\mu \alpha_1 +\nu \alpha_3}{2}$
in all cases and that in the four cases $(\epsilon ,\epsilon')$,
where each $\epsilon$ and $\epsilon'$ is $+$ or $-$ one has:
$\gamma^{\prime\natural}=-\frac{\epsilon \alpha_1+\epsilon'\alpha_3}{2}$, 
$s_{\fg} +s_{\gh}=-\frac{1}{4}\big((\mu -\epsilon 1)^2 +
      (\nu -\epsilon'1)^2\big)+\frac{1}{2}$.

Furthermore, let
\begin{gather*}
  \hat{\Delta}^{(1/2)}_{++} = \{\big( m-\frac{\mu +\nu}{2}\big)
     \delta +\alpha_2 \, , \, \big( m+\frac{\mu +\nu}{2}  \big)
      \delta +\alpha_1 +\alpha_2+\alpha_3 \, , \, \\
\big( m+\frac{\mu-\nu}{2} \big) \delta +\alpha_1 +\alpha_2\, , \,
     \big( m-\frac{\mu -\nu}{2}\big ) \delta +\alpha_2 +\alpha_3
     | m \in \ZZ_+ \}\, ,
\end{gather*}
and define $\hat{\Delta}^{(0)}_{++}$ in the four cases as follows
$(m \in \ZZ_+)$:
\begin{gather*}
  (++): \{ (m+\mu)\delta +\alpha_1 , 
      (m+\nu)\delta+\alpha_3 ,  (m+1-\mu)\delta-\alpha_1, \\
      (m+1-\nu)\delta-\alpha_3 \}\, , \\
  (+-): \{ (m+\mu )\delta +\alpha_1  , 
       (m+1+\nu)\delta +\alpha_3 , 
       (m+1-\mu)\delta -\alpha_1  , \\ 
       (m-\nu)\delta -\alpha_3 \}  , \\
  (-+): \{ (m+1+\mu)\delta +\alpha_1 ,  
        (m+\nu)\delta +\alpha_3 ,  
        (m-\mu)\delta -\alpha_1, \\ 
        (m+1-\nu)\delta -\alpha_3 \}  , \\
   (--): \{ (m+1+\mu)\delta +\alpha_1  , 
         (m+1+\nu)\delta +\alpha_3, 
         (m-\mu)\delta -\alpha_1  , \\
         (m-\nu)\delta -\alpha_3 \}\, .
\end{gather*}
Then $\hat{\Delta}^{\re}_{++} =\hat{\Delta}^{(0)}_{++} \cup
\hat{\Delta}^{(1/2)}_{++} \cup \{ m\delta +\theta |\, m \in \ZZ_+ \}$.

Next, $\fh_{W,\sigma}= \fhs \oplus\CC L^{\tw}_0$, and
$\Delta^{+,\re}_{W,\sigma} =\Delta^{+(1/2)}_{W,\sigma} \cup
\Delta^{+(0)}_{W,\sigma} \subset \fh^*_{W,\sigma}$, where
\begin{gather*}
  \Delta^{+(1/2)}_{W,\sigma} =\{ \big(m+\frac{1}{2}
      -\frac{\mu +\nu}{2}\big) \delta'-\frac{\alpha +\alpha'}{2}
      \, , \, \big( m+\frac{1}{2}+\frac{\mu+\nu}{2}\big)
      \delta'+\frac{\alpha +\alpha'}{2}\, , \, \\
      \big( m+\frac{1}{2}+\frac{\mu -\nu}{2}\big) \delta' +
      \frac{\alpha -\alpha'}{2}\, , \, \big( m+\frac{1}{2}-
      \frac{\mu -\nu}{2}\big)\delta'-\frac{\alpha -\alpha'}{2}
      |\, m \in \ZZ_+ \} \, , 
\end{gather*}
and $\Delta^{+(0)}_{W,\sigma}$ in the four cases is as follows
$(m \in \ZZ_+)$:
\begin{gather*}
  (++): \{ (m+\mu)\delta' +\alpha , 
      (m+\nu)\delta' +\alpha' ,  (m+1-\mu)\delta'-\alpha, \\
      (m+1-\nu)\delta'-\alpha' \}\, , \\
  (+-): \{ (m+\mu )\delta' +\alpha  , 
       (m+1+\nu)\delta' +\alpha' , 
       (m+1-\mu)\delta' -\alpha  , \\ 
       (m-\nu)\delta' -\alpha' \}  , \\
  (-+): \{ (m+1+\mu)\delta' +\alpha ,  
        (m+\nu)\delta' +\alpha' ,  
        (m-\mu)\delta' -\alpha , \\ 
        (m+1-\nu)\delta' -\alpha' \}  , \\
   (--): \{ (m+1+\mu)\delta' +\alpha  , 
         (m+1+\nu)\delta' +\alpha', 
         (m-\mu)\delta' -\alpha  , \\
         (m-\nu)\delta' -\alpha' \}\, .
\end{gather*}

The multiplicities of all these roots of $W(\fg ,\sigma,
\theta/2)$ are $1$.  There are, in addition, roots $m\delta'$ ($m
\in \NN$), all of multiplicity~$3$.

Let $P^{\mu ,\nu}_{N=4} (\eta)$ be the corresponding partition
function.  Let $h$, $j$ and $j'$ be the respective eigenvalues of
$L^{\tw}_0$, $J^{\tw}_0$ and $J^{\prime\tw}_0$ on $v_{\lambda}$,
so that $\lambda^{\natural} =\frac{1}{2}(j\alpha +j'\alpha')$.

Formulas (\ref{eq:4.8}) --- (\ref{eq:4.10}) give the following
expressions for the factors of the determinant in the
$(++)$~case:
\begin{eqnarray*}
  \varphi_{(m-1) \delta +\theta ,n} &=&
     h-\frac{1}{4k} (n-mk)^2 +\frac{(j+1-\mu)^2}{4k(a+1)}
       + \frac{a(j'+1-\nu)^2}{4k(a+1)}\\[1ex]
     && +\frac{k}{4}+\frac{(\mu-1)^2 +(\nu -1)^2}{4}\, ;\\
%
%
%
   \varphi_{m\delta +\beta,1} &=& h-\frac{1}{k}
       \Big(\big(  m+\frac{1}{2} \big) k+\frac{j+1-\mu}{2}
       (\beta | \alpha_1) +\frac{j'+1-\nu}{2}(\beta
       |\alpha_3)\Big)^2\\[1ex]
     &&+ \frac{(j+1-\mu)^2}{4k (a+1)} +\frac{a(j'+1-\nu)^2}
         {4k(a+1)} +\frac{k}{4} +\frac{(\mu-1)^2 +(\nu -1)^2}{4}\\[1ex]
          &&\hbox{ if }\beta \in S_{1/2}\, ;\\
   \varphi_{m\delta +\beta ,n} &=& mk +\frac{j+1-\mu}{2}
       (\beta |\alpha_1)+\frac{j'+1-\nu}{2}(\beta |\alpha_3)
       -\frac{n(\beta |\beta)}{2}\\
       && \hbox{ if }\beta \in S_0\, .
\end{eqnarray*}
The factors in the remaining three cases are obtained from the
above formulas by a shift of $\mu$ and $\nu$ as follows:
\begin{gather*}
  (-+): \mu \to \mu +2 \,, \, \nu \to \nu \, ;\\
  (+-): \mu \to \mu \, , \, \nu \to \nu +2 \, ;\\
  (--): \mu \to \mu +2 \, , \, \nu \to \nu +2 \, .
\end{gather*}

By Theorem~\ref{th:4.2} and Remark~4.2(a) we obtain the
following formula for $\det_{\eta} (k,h,j,j')$:
\begin{gather*}
 \lefteqn{\hspace{-2in} \prod_{m,n \in \NN} (k^2\varphi_{(m-1)\delta +\theta ,n}
  (h,k,j,j'))^{P^{\mu,\nu}_{N=4} (\eta -mn\delta')}}\\
\times \prod_{m\delta +\beta \in \hat{\Delta}^{(1/2)}_{++}}
   \varphi_{m\delta +\beta ,1} 
   (k,h,j,j')^{P^{\mu,\nu}_{N=4 ;(m+1/2)\delta'+\beta^{\natural}}
     (\eta -(m+1/2)\delta'-\beta^{\natural})}\\
 \lefteqn{\hspace{-2in}\times \prod_{\substack{m\delta+\beta \in
    \hat{\Delta}^{(0)}_{++}\\ n \in \NN}}
    \varphi_{m\delta +\beta ,n}
    (k,h,j,j')^{P^{\mu,\nu}_{N=4} (\eta -n
      (m\delta'+\beta^{\natural}))}\, .}
\end{gather*}

\subsection{Big $N=4$ twisted sector}\quad
\label{sec:5.7}

In this subsection we consider the involutions $\sigma =
\sigma_{\tw ,b}$ of $\fg =D(2,1;1) = osp (4,2)$ defined by:
\begin{gather*}
  \sigma (e_1)=e_3 \, , \, \sigma (e_2)=e^{-\pi i b}e_2\, , \, 
  \sigma (e_3)=e^{2\pi i b}e_1\, , \, \\
  \sigma (f_1)=f_3 \, , \, \sigma (f_2)=e^{\pi i b}f_2\, , \, 
  \sigma (f_3)=e^{-2\pi ib} f_1 \, ,
\end{gather*}
where $b \in \RR$, $-1\leq b <1$.  Introduce the following
elements of $\fg : e^{(1)}=\frac{1}{\sqrt{2}} (e_1+e^{-\pi i b}
e_3)$, $f^{(1)} = \frac{1}{\sqrt{2}} (f_1 +e^{\pi ib}f_3)$, 
$e^{(3)}=\frac{1}{\sqrt{2}}(e_1-e^{-\pi ib}e_3)$,
$f^{(3)} =\frac{1}{\sqrt{2}}(f_1-e^{\pi ib} f_3)$,
$e^{(110)}=\frac{1}{\sqrt{2}}(e_{110}+e^{-\pi i b}e_{011})$,
$f^{(110)}=\frac{1}{\sqrt{2}} (f_{110}+e^{\pi i b} f_{011})$,
$e^{(011)}=\frac{1}{\sqrt{2}}(e_{110}-e^{-\pi i b}e_{011})$, 
$f^{(011)}=\frac{1}{\sqrt{2}} (f_{110}-e^{\pi i b}f_{011})$.  We
have the following eigenspace decomposition of $\fg$ with respect
to $\sigma$ (here, as before, $\fg^{\mu}=\{ a \in \fg |\sigma (a)
=e^{2\pi i\mu}a\}$):  $\fg =\fg^0
+\fg^{1/2}+\fg^{b/2}+\fg^{-b/2}+\fg^{(1+b)/2}+\fg^{(1-b)/2}$,
where 
\begin{gather*}
\fg^0 = \hbox{span} \{ e^{(011)}, f^{(011)}, e,f,\alpha_2
,\alpha_1 +\alpha_3 \} , 
 \fg^{b/2}= \hbox{span} \{ e^{(1)},
e_{111}, f_2 \}\, ,\,\\ 
\fg^{-b/2} =\hbox{span} \{ e_2,f^{(1)}, f_{111}\},
\fg^{(1+b)/2}=\CC e^{(3)} ,\fg^{-(1+b)/2}=\CC f^{(3)}\, , \,\\
\fg^{1/2}= \hbox{span} \{ e^{(110)}, f^{(110)}, \alpha_1 -\alpha_3 \}\, .
\end{gather*}
Then $\fh^{\sigma} =\CC \theta +\CC (\alpha_1 +\alpha_3)$ and the
roots of $\hat{\fg}^{\tw}$ are described in terms of
$\tilde{\alpha}_i=\alpha_i |_{\fh^{\sigma}}$ $(i=1,2)$ and
$\delta$, the non-zero inner products between them being
$(\tilde{\alpha}_1 |\tilde{\alpha}_1)=-1/2$, $(\tilde{\alpha_1}
|\tilde{\alpha}_2)=1/2$.  The union of the above bases of the
eigenspaces of $\sigma$ is a basis of $\fg$, compatible with the
$\frac{1}{2}\ZZ$-gradation and the root space decomposition with
respect to $\fh^{\sigma}$, which we denoted by $S$. Furthermore,
$\fh^{\natural}=\CC\alpha$, where $\alpha=\alpha_1^{\natural}=-\alpha_2^{\natural}$.

We have:  $\epsilon (\sigma)=1$, $\fg_{1/2}(\sigma)_0 =\CC
e^{(110)}$, and the following possibilities for $\fn (\sigma)_{\pm}$:

\romanparenlist
\begin{enumerate}
\item 
$b \in 2\ZZ:  \fn (\sigma)_- =$ span $\{ e,e^{(110)},f^{(1)}\},
  \fn (\sigma)_+ =$ span $\{ f,f^{(110)}, e^{(1)}\}$;

\item 
$b \in 2 \ZZ +1:  \fn (\sigma)_- =$ span $\{ e,e^{(110)},
e_2,f_{111},f^{(3)}\}$,\hfill \break
\hspace*{1.4in}$\fn (\sigma)_+ =$~span $\{ f,f^{(110)},
f_2,e_{111},e^{(3)}\}$;

\item 
$b \not\in \ZZ :  \fn (\sigma)_- =\CC e+\CC e^{(110)}, \fn
(\sigma)_+ =\CC f+\CC f^{(110)}$.

\end{enumerate}

We consider separately the following two cases:
  $(+)$: $0 \leq b <1$  ; $(-)$: $-1\leq b <0$ .

The $s_i$ are as follows (in this case they depend not only on
root, but also on the root vector):
$s_e =0$, $s_{e_2}=-b /2$, $s_{e_{111}}=b /2$,
$ s_{e^{(110)}}=1/2$, $s_{e^{(011)}}=0$, $s_{e^{(3)}}=\frac{1}{2}(1+b)$, 
$s_{\alpha_1 -\alpha_3}=1/2$
in all cases; the remaining $s_i$ (up to the relation
(\ref{eq:3.3})) are:
$s_{e^{(1)}}=b /2$  in case $(+)$, $s_{e^{(1)}}=1+b /2$ in case $(-)$.
Using this, one finds that 
\begin{displaymath}
  \gamma^{\natural}_{1/2}=-\frac{b}{2}\tilde{\alpha_1}\, , \, 
  \gamma^{\prime \natural}=\mp \frac{1}{2} \tilde{\alpha}_1\,
  \hbox{and}\, s_{\fg} +s_{\gh} =-\frac{b^2}{8}\pm \frac{b}{4}
 \hbox{ in case }(\pm)\, .
\end{displaymath}

Furthermore, let $(m \in \ZZ_+)$:
\begin{displaymath}
  \hat{\Delta}^{(1/2)}_{++}=\{ \bigl( m-\frac{b}{2}\bigr)
 \delta + \tilde{\alpha}_2\, , \, \bigl( m+\frac{b}{2}\bigr)
  \delta + (2\tilde{\alpha}_1+\tilde{\alpha}_2)\}\, ,
\end{displaymath}
and define $\hat{\Delta}^{(0)}_{++}$ in cases $(\pm)$ as follows
$(m \in \ZZ_+)$:
\begin{displaymath}
  \hat{\Delta}^{(0)}_{++}= \{ \frac{m\pm b}{2}\delta \pm
  \tilde{\alpha}_1 \, , \, \frac{m+1\mp b}{2}
  \delta \mp \tilde{\alpha}_1\}\, .
\end{displaymath}
Then $\hat{\Delta}^{\re}_{++} = \hat{\Delta}^{(0)}_{++} \cup
\hat{\Delta}^{(1/2)}_{++} \cup \{ \frac{m\delta+\theta}{2}\, , \,
m\delta +\theta \,|\, m \in \ZZ_+\} $.

Next, $\fh_{W,\sigma} =\fhs \oplus \CC L^{\tw}_0$, and in cases
$(\pm)$ we have $(m \in \ZZ_+)$:
\begin{eqnarray*}
  \Delta^+_{W,\sigma}=\{ \frac{m \pm b}{2} 
     \delta' \pm \alpha\, , \, \frac{m+1\mp b}{2}\delta'
     \mp \alpha , \, (m+1)\delta' \, , \, 
     (m+\frac{1}{2})\delta' \} \\
   \cup \{ \frac{m+1}{2}\delta' \, , \, \big(
   m+\frac{1+b}{2}\big) \delta' +\alpha \, , \, 
   \big( m+\frac{1-b}{2}\big) \delta'-\alpha \}\, , 
\end{eqnarray*}
the elements from the first (resp. second) set being even
(resp. odd) roots, and the multiplicities of all roots being $1$,
except for $(m+1)\delta'$, whose multiplicity is $2$.

Let $P^b_{N=4} (\eta)$ be the corresponding partition function.
Let $h$ and $j$ be the respective eigenvalues of $L^{\tw}_0$ and
$J^{\tw ,\{ -4 \alpha_1\}}_0$, so that $\lambda^{\natural}=
\frac{j}{2}\alpha$.  Formulas~(\ref{eq:4.8}) --- (\ref{eq:4.10})
give the following expressions for the factors of the determinant
in $(\pm)$ cases:
\begin{eqnarray*}
  \varphi_{(m-1)\delta +\theta ,n} &=& h-h_{n,m}(k,j),\\
\noalign{\hbox{where}}\\
  h_{n,m} (k,j)  &=& \frac{1}{4k}\Big(\big(\frac{n}{2}-mk 
     \big)^2 -\frac{(j-b\pm 1)^2}{4}-k^2 \Big)-
     \frac{(b\mp 1)^2+1}{8}\, ;\\
\varphi_{m\delta +\beta ,1} &=& h-\frac{1}{k}
     \Big(\big(m+\frac{1}{2}\big) k+\frac{j-b\pm1}{2}
     (\beta |\tilde{\alpha}_1)\Big)^2-\frac{k^2}{4}\\
     && -\frac{(j-b\pm 1)^2}{16}  \hbox{ if }
     \beta \in S_{1/2}\backslash \{\theta/2\}\, , \\
\varphi_{m\delta +\tilde{\alpha}_1 ,n} &=&
     mk-\frac{j-b \pm 1+n}{4} \, , \, 
     \varphi_{m\delta -\tilde{\alpha}_{1},n}=
     mk +\frac{j-b\pm 1 +n}{4}\, .
\end{eqnarray*}

The extra factor is computed using formula (\ref{eq:4.7}) (in this case 
$h\spcheck_0=-1$), which gives: 
%
%
%
%
%
\begin{displaymath}
  \varphi_{(\theta -\delta)/2,0} =h+
  \frac{(j-b\pm 1)^2}{16k} +\frac{(b\mp 1)^2}{8}
  +\frac{k}{4}+\frac{1}{8}\, , 
\end{displaymath}
This again confirms our conjecture made in Remark~4.2(c).

By Theorem~\ref{th:4.2} and Remarks~4.2(a) and (b), we obtain the
following formula for $\det_{\eta}(k,h,j)$:
\begin{align*}
 \lefteqn{\hspace{.5in} \varphi_{(\theta -\delta)/2,0} 
     (k,h,j)^{P^{\prime b}_{N=4}(\eta)}\prod_{m,n \in \NN}
     k^{P^b_{N=4}(\eta -mn \delta')}}\\
&\times \prod_{\substack{m,n \in \NN \\ m+n\,\odd}}
    (h-h_{n,m} (k,j))^{P^b_{N=4}(\eta-\frac{1}{2} mn\delta')}\\
& \times \prod_{\substack{m\delta +\beta \in
    \hat{\Delta}^{(1/2)}_{++}\\ \beta =\tilde{\alpha}_2
    ,\tilde{\alpha}_1 +\tilde{\alpha}_2 , 2\tilde{\alpha}_1 
    +\tilde{\alpha}_2}} \varphi_{m\delta +\beta ,1}
    (k,h,j)^{P^b_{N=4;(m+\frac{1}{2})\delta'+\beta^{\natural}}(\eta -(m+\frac{1}{2})\delta'- \beta^{\natural})}\\
& \times \prod_{\substack{m\delta \pm \tilde{\alpha}_1\in
    \hat{\Delta}^{(0)}_{++}\\ n\in \NN}}
    \varphi_{m\delta \pm \alpha_1,n}
    (k,j)^{P^b_{N=4}(\eta -n(m\delta' \pm
      \tilde{\alpha}^{\natural}_1))}\, .
\end{align*}

\end{document}